\pdfoutput=1
\documentclass[runningheads]{llncs}

\usepackage{booktabs}   
\usepackage{subcaption} 
\usepackage{caption}
\usepackage{subcaption}
\makeatletter
\@ifundefined{lhs2tex.lhs2tex.sty.read}%
  {\@namedef{lhs2tex.lhs2tex.sty.read}{}%
   \newcommand\SkipToFmtEnd{}%
   \newcommand\EndFmtInput{}%
   \long\def\SkipToFmtEnd#1\EndFmtInput{}%
  }\SkipToFmtEnd

\newcommand\ReadOnlyOnce[1]{\@ifundefined{#1}{\@namedef{#1}{}}\SkipToFmtEnd}
\usepackage{amstext}
\usepackage{amssymb}
\usepackage{stmaryrd}
\DeclareFontFamily{OT1}{cmtex}{}
\DeclareFontShape{OT1}{cmtex}{m}{n}
  {<5><6><7><8>cmtex8
   <9>cmtex9
   <10><10.95><12><14.4><17.28><20.74><24.88>cmtex10}{}
\DeclareFontShape{OT1}{cmtex}{m}{it}
  {<-> ssub * cmtt/m/it}{}

\DeclareFontShape{OT1}{cmtt}{bx}{n}
  {<5><6><7><8>cmtt8
   <9>cmbtt9
   <10><10.95><12><14.4><17.28><20.74><24.88>cmbtt10}{}
\DeclareFontShape{OT1}{cmtex}{bx}{n}
  {<-> ssub * cmtt/bx/n}{}

\newcommand{\Conid}[1]{\mathit{#1}}
\newcommand{\Varid}[1]{\mathit{#1}}
\newcommand{\anonymous}{\kern0.06em \vbox{\hrule\@width.5em}}
\newcommand{\plus}{\mathbin{+\!\!\!+}}

\newcommand{\rbind}{\mathbin{=\mkern-6.7mu<\!\!\!<}}

\usepackage{polytable}

\@ifundefined{mathindent}%
  {\newdimen\mathindent\mathindent\leftmargini}%
  {}%

\def\resethooks{%
  \global\let\SaveRestoreHook\empty
  \global\let\ColumnHook\empty}
\newcommand*{\savecolumns}[1][default]%
  {\g@addto@macro\SaveRestoreHook{\savecolumns[#1]}}
\newcommand*{\restorecolumns}[1][default]%
  {\g@addto@macro\SaveRestoreHook{\restorecolumns[#1]}}
\newcommand*{\aligncolumn}[2]%
  {\g@addto@macro\ColumnHook{\column{#1}{#2}}}

\resethooks

\newcommand{\onelinecommentchars}{\quad-{}- }
\newcommand{\commentbeginchars}{\enskip\{-}
\newcommand{\commentendchars}{-\}\enskip}

\newcommand{\visiblecomments}{%
  \let\onelinecomment=\onelinecommentchars
  \let\commentbegin=\commentbeginchars
  \let\commentend=\commentendchars}

\newcommand{\invisiblecomments}{%
  \let\onelinecomment=\empty
  \let\commentbegin=\empty
  \let\commentend=\empty}

\visiblecomments

\newlength{\blanklineskip}
\newcommand{\hsindent}[1]{\quad}
\let\hspre\empty
\let\hspost\empty

\EndFmtInput
\makeatother
\ReadOnlyOnce{polycode.fmt}%
\makeatletter

\newcommand{\hsnewpar}[1]%
  {{\parskip=0pt\parindent=0pt\par\vskip #1\noindent}}

\newcommand{\hscodestyle}{}

\newcommand{\sethscode}[1]%
  {\expandafter\let\expandafter\hscode\csname #1\endcsname
   \expandafter\let\expandafter\endhscode\csname end#1\endcsname}

  {\par\noindent
   \advance\leftskip\mathindent
   \hscodestyle
   \let\\=\@normalcr
   \let\hspre\(\let\hspost\)%
   \pboxed}%
  {\endpboxed\)%
   \par\noindent
   \ignorespacesafterend}

  {\hsnewpar\abovedisplayskip
   \advance\leftskip\mathindent
   \hscodestyle
   \let\hspre\(\let\hspost\)%
   \pboxed}%
  {\endpboxed%
   \hsnewpar\belowdisplayskip
   \ignorespacesafterend}

  {\hsnewpar\abovedisplayskip
   \advance\leftskip\mathindent
   \hscodestyle
   \let\\=\@normalcr
   \(\pboxed}%
  {\endpboxed\)%
   \hsnewpar\belowdisplayskip
   \ignorespacesafterend}

\newcommand{\plainhs}{\sethscode{plainhscode}}

\plainhs

  {\hsnewpar\abovedisplayskip
   \advance\leftskip\mathindent
   \hscodestyle
   \let\\=\@normalcr
   \(\parray}%
  {\endparray\)%
   \hsnewpar\belowdisplayskip
   \ignorespacesafterend}

  {\parray}{\endparray}

  {\(\parray}{\endparray\)}

\def\codeframewidth{\arrayrulewidth}
\RequirePackage{calc}

  {\parskip=\abovedisplayskip\par\noindent
   \hscodestyle
   \arrayrulewidth=\codeframewidth
   \tabular{@{}|p{\linewidth-2\arraycolsep-2\arrayrulewidth-2pt}|@{}}%
   \hline\framedhslinecorrect\\{-1.5ex}%
   \let\endoflinesave=\\
   \let\\=\@normalcr
   \(\pboxed}%
  {\endpboxed\)%
   \framedhslinecorrect\endoflinesave{.5ex}\hline
   \endtabular
   \parskip=\belowdisplayskip\par\noindent
   \ignorespacesafterend}

\newcommand{\framedhslinecorrect}[2]%
  {#1[#2]}

  {\(\def\column##1##2{}%
   \let\>\undefined\let\<\undefined\let\\\undefined
   \newcommand\>[1][]{}\newcommand\<[1][]{}\newcommand\\[1][]{}%
   \def\fromto##1##2##3{##3}%
   }{\) }%

  {\let\orighscode=\hscode
   \let\origendhscode=\endhscode
   \def\endhscode{\def\hscode{\endgroup\def\@currenvir{hscode}\\}\begingroup}
   \orighscode\def\hscode{\endgroup\def\@currenvir{hscode}}}%
  {\origendhscode
   \global\let\hscode=\orighscode
   \global\let\endhscode=\origendhscode}%

\makeatother
\EndFmtInput
\usepackage{graphicx}
\usepackage{acronym}
\usepackage{tikz-cd}
\usetikzlibrary{arrows.meta}
\usetikzlibrary{decorations.markings}
\usepackage{pgfplots}
\usepackage{verbatim}
\usepackage{ marvosym }
\usepackage{listings}
\usepackage{multibib}
\usepackage{footmisc}

\usepackage{subfiles} 

\graphicspath{{../src/diagrams/}}

\ReadOnlyOnce{forall.fmt}%
\makeatletter

\let\HaskellResetHook\empty
\newcommand*{\AtHaskellReset}[1]{%
  \g@addto@macro\HaskellResetHook{#1}}
\newcommand*{\HaskellReset}{\HaskellResetHook}

\newcommand\hsforall{\global\let\hsdot=\hsperiodonce}
\newcommand*\hsperiodonce[2]{#2\global\let\hsdot=\hscompose}
\newcommand*\hscompose[2]{#1}

\AtHaskellReset{\global\let\hsdot=\hscompose}

\HaskellReset

\makeatother
\EndFmtInput

\newmuskip\codemuskip
\providecommand\codeskip{\mskip\codemuskip}%
\let\codefont\textsf

\providecommand\keyw[1]{{\codefont{\textbf{#1}}}}

\renewcommand\Varid[1]{\codefont{#1}}
\let\Conid\Varid

\definecolor{forestgreen}{rgb}{0.0, 0.27, 0.13}

\begin{document}
\title{\ensuremath{\Conid{CircuitFlow}}: A Domain Specific Language for Dataflow Programming*}
\titlerunning{\ensuremath{\Conid{CircuitFlow}}}
%
\author{Riley Evans
\and
Samantha Frohlich\orcidID{0000-0002-4423-6918}
\and
Meng Wang\orcidID{0000-0001-7780-630X}}

\authorrunning{R. Evans et al.}
%
\institute{University of Bristol, Bristol, United Kingdom
}
\maketitle              
\begin{abstract}
Dataflow applications, such as machine learning algorithms, can run for days, making it desirable to have assurances that they will work correctly.
Current tools are not good enough: too often the interactions between tasks are not type-safe, leading to undesirable runtime errors.
This paper presents a new declarative Haskell \ac{eDSL} for dataflow programming: \ensuremath{\Conid{CircuitFlow}}.
Defined as a \ac{SMP} on data that models dependencies in the workflow, it has a strong mathematical basis, refocusing on how data flows through an application, resulting in a more expressive solution that not only catches errors statically, but also achieves competitive run-time performance. In our preliminary evaluation, \ensuremath{\Conid{CircuitFlow}} outperforms the industry-leading Luigi library of Spotify by scaling better with the number of inputs.
The innovative creation of \ensuremath{\Conid{CircuitFlow}} is also of note, exemplifying how to create a modular \ac{eDSL} whose semantics necessitates effects, and where storing complex type information for program correctness is paramount.
\end{abstract}

\vspace{-2.5em}
\renewcommand{\thefootnote}{*}
\footnotetext{Published in proceedings of PADL 2022. This version includes appendices.}
\renewcommand{\thefootnote}{\arabic{footnote}}
\addtocounter{footnote}{-1}




\acrodef{DSL}{Domain Specific Language}
\acrodef{eDSL}{Embedded DSL}
\acrodef{FIFO}{First-In First-Out}
\acrodef{KPN}{Kahn Process Network}
\acrodef{GPL}{General Purpose Language}
\acrodef{DPN}{Data Process Network}
\acrodef{DAG}{Directed Acyclic Graph}
\acrodef{PID}{Process Identifier}
\acrodef{AST}{Abstract Syntax Tree}
\acrodef{SMC}{Symmetric Monoidal Category}
\acrodefplural{SMC}[SMCs]{Symmetric Monoidal Categories}
\acrodef{SMP}{Symmetric Monoidal Preorder}

\section{Introduction}


\ensuremath{\Conid{CircuitFlow}}'s domain is \emph{dataflow programming}~\cite{10.1145/642089.642111}, which deals with processing data through transformations with interlinking dependencies.
Inputs are transformed into outputs by \emph{tasks}, organised into \emph{workflows} taking the form of \acp{DAG} encoding dependencies, where the directionality indicates the direction the data is flowing, and the acyclicity ensures that the data doesn't go round in circles.
Dataflow programming is highly applicable with numerous uses spanning from scientific data analysis~\cite{dataflowforscience,dataflowforscience2} to machine learning~\cite{tensorflow,spark}.
Examples include \textit{Data Pipelines}, \textit{CI Systems}, \textit{Quartz Composer}~\cite{quartz} and \textit{Spreadsheets}.
It also has the following benefits:

\paragraph{Declarative}
Describing the shape of the DAG instead of just indicating the connections, provides a more user-friendly and declarative experience.

\paragraph{Implicit Parallelism}
Since each node in a dataflow is a pure function, it is possible to parallelise implicitly.
The purity of the nodes means that outside of data dependencies encoded in the dataflow graph, no node can interact with another.
Thus eliminating the ability for a deadlock to occur.

\paragraph{Visual}
The dataflow paradigm uses graphs. This provides the programs with a visual interpretation, allowing end-user programmer to reason visually about how data passes through the program, much easier than in an imperative approach~\cite{Hils1992VisualLA}.

Existing dataflow libraries such as Spotify's Luigi~\cite{spotify_luigi} or Apache's Airflow~\cite{airflow} have no mechanism to ensure the dependencies are valid. There is no static checking that the connections in the graph match up, which could cause runtime crashes, or even worse, the bug could go unnoticed and cause havoc in later tasks.
Consider an example shown in the docs for Luigi~\cite{spotify_luigi_docs_2020} that is made up of two tasks:
the first, \texttt{GenerateWords}, generates a list of words and saves it to a file; and the second, \texttt{CountLetters}, counts the number of letters in each of those words.
An implementation of this in Luigi could have a very subtle bug! \texttt{GenerateWords} could write the words to a file separated by new lines, while \texttt{CountLetters} expects a comma-separated list.
This shows a key flaw in this system, as it is up to the programmer to ensure that they write the outputs correctly,
and then that they read that same file in the same way.
This error, would not even cause a run-time error, instead, it will just produce the incorrect result.
For a developer, this is extremely unhelpful: it means more time is used writing tests --- something that no one enjoys.
With good development practices, the risk is reduced, but as functional programmers, we know a better way: abstraction and static typing.

Why not eliminate all of this with an abstraction of the reading and writing of many different sources and types?
The abstraction will help to ensure correctness of passing data via files by eliminating any possible duplicated code.
Instead, just having a uniform interface to test.
Then the abstract interface can be combined with the type system so that in each program, it is enforced that the types align.

This promotes the need for a new solution with such features that can safely compose tasks and make use of types to perform static analysis to ensure that dependencies are valid.

We present \ensuremath{\Conid{CircuitFlow}}, which takes a different line of attack from its predecessing plumbers like Luigi: rather than focus on how to compose tasks together, it defines a declarative language that describes how data flows through a workflow.
In \ensuremath{\Conid{CircuitFlow}}, it would not be possible to feed the output of one task, with the type \ensuremath{\Conid{FileStore}\codeskip [\mskip1.5mu \Conid{String}\mskip1.5mu]} into a task that expects a \ensuremath{\Conid{CommaSepFile}\codeskip [\mskip1.5mu \Conid{String}\mskip1.5mu]}.
The same example, written in \ensuremath{\Conid{CircuitFlow}}, is defined as:

\begin{hscode}\SaveRestoreHook
\column{B}{@{}>{\hspre}l<{\hspost}@{}}%
\column{3}{@{}>{\hspre}l<{\hspost}@{}}%
\column{5}{@{}>{\hspre}l<{\hspost}@{}}%
\column{21}{@{}>{\hspre}l<{\hspost}@{}}%
\column{26}{@{}>{\hspre}l<{\hspost}@{}}%
\column{E}{@{}>{\hspre}l<{\hspost}@{}}%
\>[B]{}\Varid{generateWords}\mathbin{::}\Conid{Circuit}\codeskip '[\mskip1.5mu \Conid{Var}\mskip1.5mu]\codeskip '[\mskip1.5mu ()\mskip1.5mu]\codeskip '[\mskip1.5mu \Conid{FileStore}\mskip1.5mu]\codeskip '[\mskip1.5mu [\mskip1.5mu \Conid{String}\mskip1.5mu]\mskip1.5mu]\codeskip \Conid{N1}{}\<[E]%
\\
\>[B]{}\Varid{generateWords}\mathrel{=}\Varid{functionTask}\codeskip (\Varid{const}\codeskip [\mskip1.5mu \text{\ttfamily \char34 apple\char34},\text{\ttfamily \char34 banana\char34},\text{\ttfamily \char34 grapefruit\char34}\mskip1.5mu]){}\<[E]%
\\[\blanklineskip]%
\>[B]{}\Varid{countLetters}\mathbin{::}\Conid{Circuit}\codeskip {}\<[26]%
\>[26]{}'[\mskip1.5mu \Conid{CommaSepFile}\mskip1.5mu]\codeskip '[\mskip1.5mu [\mskip1.5mu \Conid{String}\mskip1.5mu]\mskip1.5mu]\codeskip '[\mskip1.5mu \Conid{FileStore}\mskip1.5mu]\codeskip '[\mskip1.5mu [\mskip1.5mu \Conid{String}\mskip1.5mu]\mskip1.5mu]\codeskip \Conid{N1}{}\<[E]%
\\
\>[B]{}\Varid{countLetters}\mathrel{=}\Varid{functionTask}\codeskip (\Varid{map}\codeskip \Varid{f}){}\<[E]%
\\
\>[B]{}\hsindent{3}{}\<[3]%
\>[3]{}\mathbf{where}{}\<[E]%
\\
\>[3]{}\hsindent{2}{}\<[5]%
\>[5]{}\Varid{f}\codeskip \Varid{word}\mathrel{=}(\Varid{concat}\codeskip [\mskip1.5mu \Varid{word},\text{\ttfamily \char34 :\char34},\Varid{show}\codeskip (\Varid{length}\codeskip \Varid{word})\mskip1.5mu]){}\<[E]%
\\[\blanklineskip]%
\>[B]{}\Varid{circuit}\mathbin{::}\Conid{Circuit}\codeskip {}\<[21]%
\>[21]{}'[\mskip1.5mu \Conid{Var}\mskip1.5mu]\codeskip '[\mskip1.5mu ()\mskip1.5mu]\codeskip '[\mskip1.5mu \Conid{FileStore}\mskip1.5mu]\codeskip '[\mskip1.5mu [\mskip1.5mu \Conid{String}\mskip1.5mu]\mskip1.5mu]\codeskip \Conid{N1}{}\<[E]%
\\
\>[B]{}\Varid{circuit}\mathrel{=}\Varid{generateWords}<\!\!\!\!-\!\!\!\!>\Varid{countLetters}{}\<[E]%
\ColumnHook
\end{hscode}\resethooks

\noindent
In this example, it will fail to compile, giving the error:

\text{\ttfamily \char62{}~Couldn\char39{}t~match~type~\char96{}}\ensuremath{\Conid{CommaSepFile}}\text{\ttfamily \char39{}~with~\char96{}}\ensuremath{\Conid{FileStore}}\text{\ttfamily \char39{}}

\noindent
Benefiting the user since the feedback loop of knowing if the program will succeed is reduced.
Previously, the whole data pipeline had to be run, whereas now this information is available at compile-time.

Due to the type heft required for such a language, which includes DataKinds~\cite{10.1145/2103786.2103795}, Singletons~\cite{10.1145/2364506.2364522}, Type Families~\cite{10.1145/1411204.1411215}, Heterogeneous lists~\cite{10.1145/1017472.1017488}, Phantom Types (a brief introduction of which can be found in Appendix A), it will be embedded.

\ensuremath{\Conid{CircuitFlow}} draws its origins from monoidal resource theory~\cite{Coecke_2016}, details of which can be found in Appendix~ \ref{app:resource-theories}. It is then compiled down to a \acf{KPN} that executes the workflow in parallel, to provide the speed benefits of multi-core processors.
The \ac{KPN} used by \ensuremath{\Conid{CircuitFlow}} is capable of handling an exception in a task, without causing the full network to crash, allowing computation to continue after for successive inputs.

\subsubsection{Contributions:}
A declarative \ac{eDSL} for creating dataflow programs that:
  \begin{itemize}
    \item employs state of the art DSL design techniques, including indexed data types \`{a} la carte and principled recursion to provide interpretations for the \acs{AST}.
    \item uses state of the art Haskell methods to produce a type-safe implementation.
    \item makes use of indexed functors, extended to support multiple indicies, to construct a type-indexed \acs{AST} in conjunction with an indexed monadic catamorphism to provide a type-safe translation to a \acs{KPN}.
    \item has a strong mathematical grounding in monadic resource theories providing confidence that the language can represent all dataflow diagrams.
    \item has appealing preliminary benchmark performance against another competing library --- outperforming Luigi by almost 4x on large numbers of inputs.
    \item exemplifies how to create such a language in a modular manner.
    \item uses the first known implementation of a \acl{KPN} in Haskell.
  \end{itemize}
Examples that demonstrate the language's applicability:
  \begin{itemize}
        \item Machine learning: preprocessing of real world song data in comparison to Spotify's Luigi.
        \item Build systems: the thesis this paper is based on was compiled using \ensuremath{\Conid{CircuitFlow}} (details in Appendix B).
  \end{itemize}
\section{\ensuremath{\Conid{CircuitFlow}} Language}\label{sec:lang}

\begin{figure}[ht]
  \centering
  \begin{tikzpicture}[node distance=3cm, main/.style = {draw, thick}]
    \node (Month1)                   {Month 1};
    \node (Month2) [below of=Month1, yshift=25mm] {Month 2};
    \node (Month3) [below of=Month2, yshift=25mm] {Month 3};

    \node[main] (AggSongs)    [right of=Month1] {AggSongs};
    \node[main] (AggArtists)  [right of=Month3] {AggArtists};

    \node[main] (Top10a)  [right of=AggSongs]   {Top10};
    \node[main] (Top10b)  [right of=AggArtists] {Top10};
    \node (Top10Songs)   [right of=Top10a] {T10 Songs};
    \node (Top10Artists) [right of=Top10b] {T10 Artists};
    \draw[->, >=stealth] (Month1)     -- (AggSongs);
    \draw[->, >=stealth] (Month2)     -- (AggSongs);
    \draw[->, >=stealth] (Month3)     -- (AggSongs);
    \draw[->, >=stealth] (Month1)     -- (AggArtists);
    \draw[->, >=stealth] (Month2)     -- (AggArtists);
    \draw[->, >=stealth] (Month3)     -- (AggArtists);
    \draw[->, >=stealth] (AggSongs)   -- (Top10a);
    \draw[->, >=stealth] (AggArtists) -- (Top10b);
    \draw[->, >=stealth] (Top10a)     -- (Top10Songs);
    \draw[->, >=stealth] (Top10b)     -- (Top10Artists);
  \end{tikzpicture}
  \caption{A dataflow diagram for pre-processing the song data}
  \label{fig:example-song-pre-proc}
\end{figure}
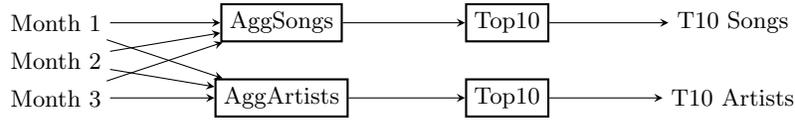

A use case for \ensuremath{\Conid{CircuitFlow}} is building data pipelines for machine learning. Consider the example where an audio streaming service would like to create a playlist full of new songs to listen to. This could require a machine learning model that can predict songs based on the top ten artists and songs that the user has listened to over the last three months. However, each of the months' data is stored in different files that need aggregating together before they can be input into the model. This problem can be drawn up as a dataflow diagram like Figure~\ref{fig:example-song-pre-proc}. To achieve this preprocessing, a software developer at said audio streaming service would need to use the following key features of the \ensuremath{\Conid{CircuitFlow}} language.

\subsection{DataStores}

Dataflow programming revolves around transforming inputs into outputs. Thus the first thing the language needs is a way of getting inputs and writing outputs. For the preprocessing example, this corresponds to a way of interfacing with the different months of song data; a way to pass on the aggregated songs and artists to the top ten calculators; and finally somewhere to store the preprocessed output ready for the machine learning model.
In \ensuremath{\Conid{CircuitFlow}}, \ensuremath{\Conid{DataStore}}s are used to pass values between different tasks, in a closely controlled manner. To abstract over the different ways of storing data, they are defined as a type class:
\begin{hscode}\SaveRestoreHook
\column{B}{@{}>{\hspre}l<{\hspost}@{}}%
\column{3}{@{}>{\hspre}l<{\hspost}@{}}%
\column{5}{@{}>{\hspre}l<{\hspost}@{}}%
\column{12}{@{}>{\hspre}l<{\hspost}@{}}%
\column{20}{@{}>{\hspre}c<{\hspost}@{}}%
\column{20E}{@{}l@{}}%
\column{24}{@{}>{\hspre}l<{\hspost}@{}}%
\column{E}{@{}>{\hspre}l<{\hspost}@{}}%
\>[3]{}\mathbf{class}\codeskip \Conid{DataStore}\codeskip \Varid{f}\codeskip \Varid{a}\codeskip \mathbf{where}{}\<[E]%
\\
\>[3]{}\hsindent{2}{}\<[5]%
\>[5]{}\Varid{fetch}{}\<[12]%
\>[12]{}\mathbin{::}\Varid{f}\codeskip \Varid{a}{}\<[20]%
\>[20]{}\to {}\<[20E]%
\>[24]{}\Conid{IO}\codeskip \Varid{a}{}\<[E]%
\\
\>[3]{}\hsindent{2}{}\<[5]%
\>[5]{}\Varid{save}{}\<[12]%
\>[12]{}\mathbin{::}\Varid{f}\codeskip \Varid{a}{}\<[20]%
\>[20]{}\to {}\<[20E]%
\>[24]{}\Varid{a}\to \Conid{IO}\codeskip (){}\<[E]%
\\
\>[3]{}\hsindent{2}{}\<[5]%
\>[5]{}\Varid{empty}{}\<[12]%
\>[12]{}\mathbin{::}\Conid{TaskUUID}\to \Conid{JobUUID}\to \Conid{IO}\codeskip (\Varid{f}\codeskip \Varid{a}){}\<[E]%
\ColumnHook
\end{hscode}\resethooks
The type class provides a way of extracting a value from a \ensuremath{\Conid{DataStore}} (\ensuremath{\Varid{fetch}}), a way to write to one (\ensuremath{\Varid{save}}), and a way of creating an empty one for a specific task.
Although the user can define their own, the library comes with  predefined \ensuremath{\Conid{DataStore}}s, the simplest is a \ensuremath{\Conid{Var}}, based on \ensuremath{\Conid{MVar}}s (mutable locations).
\begin{hscode}\SaveRestoreHook
\column{B}{@{}>{\hspre}l<{\hspost}@{}}%
\column{3}{@{}>{\hspre}l<{\hspost}@{}}%
\column{5}{@{}>{\hspre}l<{\hspost}@{}}%
\column{E}{@{}>{\hspre}l<{\hspost}@{}}%
\>[3]{}\mathbf{newtype}\codeskip \Conid{Var}\codeskip \Varid{a}\mathrel{=}\Conid{Var}\codeskip \{\mskip1.5mu \Varid{unVar}\mathbin{::}\Conid{MVar}\codeskip \Varid{a}\mskip1.5mu\}\codeskip \mathbf{deriving}\codeskip (\Conid{Eq}){}\<[E]%
\\[\blanklineskip]%
\>[3]{}\mathbf{instance}\codeskip \Conid{DataStore}\codeskip \Conid{Var}\codeskip \Varid{a}\codeskip \mathbf{where}{}\<[E]%
\\
\>[3]{}\hsindent{2}{}\<[5]%
\>[5]{}\Varid{fetch}\mathrel{=}\Varid{readMVar}\hsdot{\cdot }{.}\Varid{unVar}{}\<[E]%
\\
\>[3]{}\hsindent{2}{}\<[5]%
\>[5]{}\Varid{save}\mathrel{=}\Varid{putMVar}\hsdot{\cdot }{.}\Varid{unVar}{}\<[E]%
\\
\>[3]{}\hsindent{2}{}\<[5]%
\>[5]{}\Varid{empty}\codeskip \anonymous \codeskip \anonymous \mathrel{=}\Conid{Var}<\!\!\$\!\!>\Varid{newEmptyMVar}{}\<[E]%
\ColumnHook
\end{hscode}\resethooks
\ensuremath{\Conid{Var}} doesn't use its \ensuremath{\Varid{id}} arguments in \ensuremath{\Varid{empty}}, however, other predefined stores, such as \ensuremath{\Conid{FileStore}} and \ensuremath{\Conid{CSVStore}}, use them to decide where to place the files created.

\paragraph{Combined DataStores}
A special case of a \ensuremath{\Conid{DataStore}}, they allow the interfacing with typed lists, not just a single type.
The typed list is a variation on \ensuremath{\Conid{HList}}s: \ensuremath{\Conid{IHList}} (defined below).
Combined data stores are automatically derived from existing \ensuremath{\Conid{DataStore}} instances, making it easier for tasks to fetch from multiple inputs by supplying \ensuremath{\Varid{fetch'}}.
(Since tasks can only have one output, there is no need for a \ensuremath{\Varid{save'}} function.)
\begin{hscode}\SaveRestoreHook
\column{B}{@{}>{\hspre}l<{\hspost}@{}}%
\column{3}{@{}>{\hspre}l<{\hspost}@{}}%
\column{5}{@{}>{\hspre}l<{\hspost}@{}}%
\column{13}{@{}>{\hspre}l<{\hspost}@{}}%
\column{33}{@{}>{\hspre}l<{\hspost}@{}}%
\column{E}{@{}>{\hspre}l<{\hspost}@{}}%
\>[3]{}\mathbf{data}\codeskip \Conid{IHList}\codeskip (\Varid{fs}\mathbin{::}[\mskip1.5mu \mathbin{*}\to \mathbin{*}\mskip1.5mu])\codeskip (\Varid{as}\mathbin{::}[\mskip1.5mu \mathbin{*}\mskip1.5mu])\codeskip \mathbf{where}{}\<[E]%
\\
\>[3]{}\hsindent{2}{}\<[5]%
\>[5]{}\Conid{HCons'}{}\<[13]%
\>[13]{}\mathbin{::}\Varid{f}\codeskip \Varid{a}\to \Conid{IHList}\codeskip \Varid{fs}\codeskip \Varid{as}\to \Conid{IHList}\codeskip (\Varid{f}\codeskip \,'\!\!\mathbin{:}\codeskip \Varid{fs})\codeskip (\Varid{a}\codeskip \,'\!\!\mathbin{:}\codeskip \Varid{as}){}\<[E]%
\\
\>[3]{}\hsindent{2}{}\<[5]%
\>[5]{}\Conid{HNil'}{}\<[13]%
\>[13]{}\mathbin{::}\Conid{IHList}\codeskip '[\mskip1.5mu \mskip1.5mu]\codeskip '[\mskip1.5mu \mskip1.5mu]{}\<[E]%
\\[\blanklineskip]%
\>[3]{}\mathbf{class}\codeskip \Conid{DataStore}'\codeskip (\Varid{fs}\mathbin{::}[\mskip1.5mu \mathbin{*}\to \mathbin{*}\mskip1.5mu])\codeskip (\Varid{as}\mathbin{::}[\mskip1.5mu \mathbin{*}\mskip1.5mu])\codeskip \mathbf{where}{}\<[E]%
\\
\>[3]{}\hsindent{2}{}\<[5]%
\>[5]{}\Varid{fetch'}{}\<[13]%
\>[13]{}\mathbin{::}\Conid{IHList}\codeskip \Varid{fs}\codeskip \Varid{as}\to {}\<[33]%
\>[33]{}\Conid{IO}\codeskip (\Conid{HList}\codeskip \Varid{as}){}\<[E]%
\\
\>[3]{}\hsindent{2}{}\<[5]%
\>[5]{}\Varid{empty'}{}\<[13]%
\>[13]{}\mathbin{::}\Conid{TaskUUID}\to \Conid{JobUUID}\to \Conid{IO}\codeskip (\Conid{IHList}\codeskip \Varid{fs}\codeskip \Varid{as}){}\<[E]%
\ColumnHook
\end{hscode}\resethooks
\subsection{Circuit Type}
A \ensuremath{\Conid{Circuit}} represents some computation that has some number of inputs and outputs.
In order to statically check dependencies, the \ensuremath{\Conid{Circuit}} type needs to store a lot of information.
\begin{hscode}\SaveRestoreHook
\column{B}{@{}>{\hspre}l<{\hspost}@{}}%
\column{3}{@{}>{\hspre}l<{\hspost}@{}}%
\column{12}{@{}>{\hspre}l<{\hspost}@{}}%
\column{33}{@{}>{\hspre}c<{\hspost}@{}}%
\column{33E}{@{}l@{}}%
\column{37}{@{}>{\hspre}l<{\hspost}@{}}%
\column{48}{@{}>{\hspre}l<{\hspost}@{}}%
\column{60}{@{}>{\hspre}c<{\hspost}@{}}%
\column{60E}{@{}l@{}}%
\column{64}{@{}>{\hspre}l<{\hspost}@{}}%
\column{79}{@{}>{\hspre}c<{\hspost}@{}}%
\column{79E}{@{}l@{}}%
\column{E}{@{}>{\hspre}l<{\hspost}@{}}%
\>[3]{}\Conid{Circuit}\codeskip {}\<[12]%
\>[12]{}(\Varid{insContainerTypes}{}\<[33]%
\>[33]{}\mathbin{::}{}\<[33E]%
\>[37]{}[\mskip1.5mu \mathbin{*}\to \mathbin{*}\mskip1.5mu])\codeskip {}\<[48]%
\>[48]{}(\Varid{insTypes}{}\<[60]%
\>[60]{}\mathbin{::}{}\<[60E]%
\>[64]{}[\mskip1.5mu \mathbin{*}\mskip1.5mu])\codeskip {}\<[E]%
\\
\>[12]{}(\Varid{outsContainerTypes}{}\<[33]%
\>[33]{}\mathbin{::}{}\<[33E]%
\>[37]{}[\mskip1.5mu \mathbin{*}\to \mathbin{*}\mskip1.5mu])\codeskip {}\<[48]%
\>[48]{}(\Varid{outsTypes}{}\<[60]%
\>[60]{}\mathbin{::}{}\<[60E]%
\>[64]{}[\mskip1.5mu \mathbin{*}\mskip1.5mu])\codeskip (\Varid{nIns}\mathbin{::}{}\<[79]%
\>[79]{}\Conid{Nat}){}\<[79E]%
\ColumnHook
\end{hscode}\resethooks
It has five type parameters:
\ensuremath{\Varid{insContainerTypes}}, a type-list of storage types, for example \ensuremath{'[\mskip1.5mu \Conid{VariableStore},\Conid{CSVStore}\mskip1.5mu]};
\ensuremath{\Varid{insTypes}}, a type-list of the types stored in the storage, for example \ensuremath{'[\mskip1.5mu \Conid{Int},[\mskip1.5mu (\Conid{String},\Conid{Float})\mskip1.5mu]\mskip1.5mu]};
\ensuremath{\Varid{outsContainerTypes}} and \ensuremath{\Varid{outsTypes}} mirror that the examples above, but for the outputs instead.
The container and value types are separate, due to the need for them to be ``unapplied'' for the \ensuremath{\Conid{DataStore}} typeclass.
Unfortunately, GHC requires a little more information to perform this match check, such as the seemingly superfluous \ensuremath{\Varid{nIns}}, a type-level Nat that is the length of the input lists.

\subsection{Circuit Constructors}
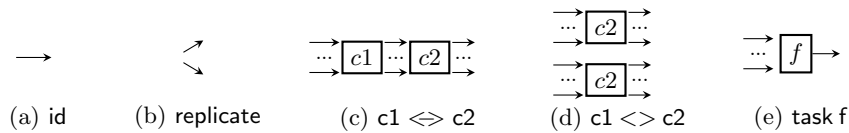
\label{sec:lang-circuit-constructors}

\begin{figure}[hbt]
\abovecaptionskip
\belowcaptionskip
\centering
\begin{subfigure}{0.14\columnwidth}
\centering
\abovecaptionskip
\belowcaptionskip
\vspace{5mm}
\begin{tikzpicture}[node distance={7mm}, main/.style = {draw, circle, thick}]
  \node (X) {$$};
  \node (Y) [right of=X] {$$};
  \draw[->, >=stealth] (X) -- (Y);
\end{tikzpicture}
\vspace{5mm}
\caption{\ensuremath{\Varid{id}}}
\end{subfigure}
\begin{subfigure}{0.19\columnwidth}
\centering
\abovecaptionskip
\belowcaptionskip
\vspace{2mm}
\begin{tikzpicture}[node distance={7mm}, main/.style = {draw, circle, thick}]
  \node (X) {$$};
  \node (Y) [above right of=X, yshift=-2mm] {$$};
  \node (Z) [below right of=X, yshift=2mm] {$$};
  \draw[->, >=stealth] (X) -- (Y);
  \draw[->, >=stealth] (X) -- (Z);
\end{tikzpicture}
\vspace{2mm}
\caption{\ensuremath{\Varid{replicate}}}
\end{subfigure}
\begin{subfigure}{0.25\columnwidth}
\centering
\abovecaptionskip
\belowcaptionskip
\vspace{4mm}
\begin{tikzpicture}[node distance={7mm}, main/.style = {draw, thick}]
  \node (X) {$$};
  \node[main] (task) [right of=X] {$c1$};
  \node[main] (task2) [right of=task, xshift=2mm] {$c2$};
  \node (Z) [right of=task2, xshift=2mm] {$$};
  \node (hidden1) [above of=X, yshift=-7mm, xshift=2mm] {\tiny\textbf{...}};
  \node (hidden2) [right of=task, xshift=-2.5mm] {\tiny\textbf{...}};
  \node (hidden3) [right of=task2, xshift=-2.5mm] {\tiny\textbf{...}};
  \draw[->, >=stealth] ([yshift=2mm]X.center) --  ([yshift=2mm,  xshift=-3mm]task.center);
  \draw[->, >=stealth] ([yshift=-2mm]X.center) -- ([yshift=-2mm, xshift=-3mm]task.center);
  \draw[->, >=stealth] ([yshift=2mm,  xshift=3mm]task.center) --  ([yshift=2mm,  xshift=-3mm]task2.center);
  \draw[->, >=stealth] ([yshift=-2mm, xshift=3mm]task.center) --  ([yshift=-2mm, xshift=-3mm]task2.center);
  \draw[->, >=stealth] ([yshift=2mm,  xshift=3mm]task2.center) -- ([yshift=2mm,  xshift=-3mm]Z.center);
  \draw[->, >=stealth] ([yshift=-2mm, xshift=3mm]task2.center) -- ([yshift=-2mm, xshift=-3mm]Z.center);
\end{tikzpicture}
\vspace{4mm}
\caption{\ensuremath{\Varid{c1}<\!\!\!\!-\!\!\!\!>\Varid{c2}}}
\end{subfigure}
\begin{subfigure}{0.19\columnwidth}
\centering
\abovecaptionskip
\belowcaptionskip
\begin{tikzpicture}[node distance={7mm}, main/.style = {draw, thick}]
  \node (X) {$$};
  \node[main] (task) [right of=X] {$c2$};
  \node (Y) [right of=task, xshift=2mm] {$$};
  \node (X') [below of=X] {$$};
  \node[main] (task') [right of=X'] {$c2$};
  \node (Y') [right of=task', xshift=2mm] {$$};
  \node (hidden2) [below of=X, yshift=7mm, xshift=2mm] {\tiny\textbf{...}};
  \node (hidden2') [below of=X', yshift=7mm, xshift=2mm] {\tiny\textbf{...}};
  \node (hidden3) [right of=task, xshift=-2.5mm] {\tiny\textbf{...}};
  \node (hidden3') [right of=task', xshift=-2.5mm] {\tiny\textbf{...}};
  \draw[->, >=stealth] ([yshift=2mm]X.center) -- ([yshift=2mm, xshift=-3mm]task.center);
  \draw[->, >=stealth] ([yshift=-2mm]X.center) -- ([yshift=-2mm, xshift=-3mm]task.center);
  \draw[->, >=stealth] ([yshift=2mm, xshift=3mm]task.center) -- ([yshift=2mm, xshift=-3mm]Y.center);
  \draw[->, >=stealth] ([yshift=-2mm, xshift=3mm]task.center) -- ([yshift=-2mm, xshift=-3mm]Y.center);
  \draw[->, >=stealth] ([yshift=2mm]X'.center) -- ([yshift=2mm, xshift=-3mm]task'.center);
  \draw[->, >=stealth] ([yshift=-2mm]X'.center) -- ([yshift=-2mm, xshift=-3mm]task'.center);
  \draw[->, >=stealth] ([yshift=2mm, xshift=3mm]task'.center) -- ([yshift=2mm, xshift=-3mm]Y'.center);
  \draw[->, >=stealth] ([yshift=-2mm, xshift=3mm]task'.center) -- ([yshift=-2mm, xshift=-3mm]Y'.center);
\end{tikzpicture}
\vspace{1mm}
\caption{\ensuremath{\Varid{c1}\mathbin{<>}\Varid{c2}}}
\end{subfigure}
\begin{subfigure}{0.19\columnwidth}
\centering
\abovecaptionskip
\belowcaptionskip
\vspace{3mm}
\begin{tikzpicture}[node distance={7mm}, main/.style = {draw, thick}]
  \node (X) {$$};
  \node[main] (task) [right of=X] {$f$};
  \node (Y) [right of=task] {$$};
  \node (hidden2) [above of=X, yshift=-7mm, xshift=2mm] {\tiny\textbf{...}};
  \draw[->, >=stealth] ([yshift=2mm]X.center) -- ([yshift=2mm, xshift=-3mm]task.center);
  \draw[->, >=stealth] ([yshift=-2mm]X.center) -- ([yshift=-2mm, xshift=-3mm]task.center);
  \draw[->, >=stealth] (task) -- (Y);
\end{tikzpicture}
\vspace{4mm}
\caption{\ensuremath{\Varid{task}\codeskip \Varid{f}}}
\end{subfigure}

\label{fig:circuit-constructors}
\end{figure}

Above shows the core constructors of the language along with their diagrammatic representation.
Here the relation to resource theories is apparent,
the constructors in this library make up a \ac{SMP}, establishing them as a resource theory able to represent any \ac{DAG}.
More details can be found in appendices \ref{app:resource-theories} and \ref{app:SMP-proof}.
The diagrammatic interpretation also makes translation from dataflow diagrams, such as Figure~\ref{fig:example-song-pre-proc}, to \ensuremath{\Conid{CircuitFlow}} code easy.

In the language, there are two types of constructors: those that create basic circuits and those that compose them.
The behaviour of the constructor is recorded within the types.
Here are the types of some basic circuits:
\begin{hscode}\SaveRestoreHook
\column{B}{@{}>{\hspre}l<{\hspost}@{}}%
\column{3}{@{}>{\hspre}l<{\hspost}@{}}%
\column{14}{@{}>{\hspre}c<{\hspost}@{}}%
\column{14E}{@{}l@{}}%
\column{18}{@{}>{\hspre}l<{\hspost}@{}}%
\column{41}{@{}>{\hspre}l<{\hspost}@{}}%
\column{54}{@{}>{\hspre}c<{\hspost}@{}}%
\column{54E}{@{}l@{}}%
\column{58}{@{}>{\hspre}l<{\hspost}@{}}%
\column{67}{@{}>{\hspre}l<{\hspost}@{}}%
\column{80}{@{}>{\hspre}l<{\hspost}@{}}%
\column{93}{@{}>{\hspre}l<{\hspost}@{}}%
\column{106}{@{}>{\hspre}l<{\hspost}@{}}%
\column{119}{@{}>{\hspre}l<{\hspost}@{}}%
\column{E}{@{}>{\hspre}l<{\hspost}@{}}%
\>[3]{}\Varid{id}{}\<[14]%
\>[14]{}\mathbin{::}{}\<[14E]%
\>[18]{}\Conid{DataStore}'\codeskip '[\mskip1.5mu \Varid{f}\mskip1.5mu]\codeskip {}\<[41]%
\>[41]{}'[\mskip1.5mu \Varid{a}\mskip1.5mu]{}\<[54]%
\>[54]{}\Rightarrow {}\<[54E]%
\>[58]{}\Conid{Circuit}\codeskip {}\<[67]%
\>[67]{}'[\mskip1.5mu \Varid{f}\mskip1.5mu]\codeskip {}\<[80]%
\>[80]{}'[\mskip1.5mu \Varid{a}\mskip1.5mu]\codeskip {}\<[93]%
\>[93]{}'[\mskip1.5mu \Varid{f}\mskip1.5mu]\codeskip {}\<[106]%
\>[106]{}'[\mskip1.5mu \Varid{a}\mskip1.5mu]\codeskip {}\<[119]%
\>[119]{}\Conid{N1}{}\<[E]%
\\
\>[3]{}\Varid{replicate}{}\<[14]%
\>[14]{}\mathbin{::}{}\<[14E]%
\>[18]{}\Conid{DataStore}'\codeskip '[\mskip1.5mu \Varid{f}\mskip1.5mu]\codeskip {}\<[41]%
\>[41]{}'[\mskip1.5mu \Varid{a}\mskip1.5mu]{}\<[54]%
\>[54]{}\Rightarrow {}\<[54E]%
\>[58]{}\Conid{Circuit}\codeskip {}\<[67]%
\>[67]{}'[\mskip1.5mu \Varid{f}\mskip1.5mu]\codeskip {}\<[80]%
\>[80]{}'[\mskip1.5mu \Varid{a}\mskip1.5mu]\codeskip {}\<[93]%
\>[93]{}'[\mskip1.5mu \Varid{f},\Varid{f}\mskip1.5mu]\codeskip {}\<[106]%
\>[106]{}'[\mskip1.5mu \Varid{a},\Varid{a}\mskip1.5mu]\codeskip {}\<[119]%
\>[119]{}\Conid{N1}{}\<[E]%
\ColumnHook
\end{hscode}\resethooks
Consider the \ensuremath{\Varid{id}} constructor, for convenience the \ensuremath{\Varid{nins}} parameter is shorted with type synonyms, e.g. \ensuremath{\Conid{N1}\mathord{\sim}'\Conid{Succ}\codeskip '\Conid{Zero}}.
It can be seen how the type information for this constructor states that it has 1 input value of type \ensuremath{\Varid{f}\codeskip \Varid{a}} and it returns that same value.
Each type parameter in \ensuremath{\Varid{id}} is a phantom type, since there are no values stored in the data type that use the type parameters.
The \ensuremath{\Varid{replicate}} constructor states that a single input value of type \ensuremath{\Varid{f}\codeskip \Varid{a}} should be input, and that value should then be duplicated and output.
There is also a \ensuremath{\Varid{swap}} constructor that takes two values as input and swaps their order, and
\ensuremath{\Varid{dropL}} / \ensuremath{\Varid{dropR}} constructors that will take two inputs and drop the left or the right one respectively.

To use these basic circuits, \ensuremath{\Conid{CircuitFlow}} provides two constructors named `beside' and `then' to compose circuits.
The definition of these constructors will require type level calculations. This is where closed type families~\cite{10.1145/2535838.2535856} come in, allowing for type level versions of \ensuremath{(\mathbin{+})} and \ensuremath{(\plus )} \cite{10.1145/1017472.1017488} (requiring \ensuremath{\Conid{PolyKinds}}~\cite{10.1145/2103786.2103795}).

\paragraph{The `Then' Constructor},
denoted by \ensuremath{<\!\!\!\!-\!\!\!\!>}, is used run one circuit, \textit{then} another, encapsulating the idea of dependencies.
Through types, it enforces that the output of the first circuit is the same as the input to the second circuit.
\begin{hscode}\SaveRestoreHook
\column{B}{@{}>{\hspre}l<{\hspost}@{}}%
\column{3}{@{}>{\hspre}l<{\hspost}@{}}%
\column{5}{@{}>{\hspre}l<{\hspost}@{}}%
\column{17}{@{}>{\hspre}l<{\hspost}@{}}%
\column{21}{@{}>{\hspre}l<{\hspost}@{}}%
\column{25}{@{}>{\hspre}l<{\hspost}@{}}%
\column{29}{@{}>{\hspre}l<{\hspost}@{}}%
\column{33}{@{}>{\hspre}l<{\hspost}@{}}%
\column{49}{@{}>{\hspre}l<{\hspost}@{}}%
\column{53}{@{}>{\hspre}l<{\hspost}@{}}%
\column{57}{@{}>{\hspre}l<{\hspost}@{}}%
\column{61}{@{}>{\hspre}l<{\hspost}@{}}%
\column{65}{@{}>{\hspre}l<{\hspost}@{}}%
\column{81}{@{}>{\hspre}l<{\hspost}@{}}%
\column{85}{@{}>{\hspre}l<{\hspost}@{}}%
\column{89}{@{}>{\hspre}l<{\hspost}@{}}%
\column{93}{@{}>{\hspre}l<{\hspost}@{}}%
\column{97}{@{}>{\hspre}l<{\hspost}@{}}%
\column{E}{@{}>{\hspre}l<{\hspost}@{}}%
\>[3]{}(<\!\!\!\!-\!\!\!\!>)\mathbin{::}(\Conid{DataStore}'\codeskip \Varid{fs}\codeskip \Varid{as},\Conid{DataStore}'\codeskip \Varid{gs}\codeskip \Varid{bs},\Conid{DataStore}'\codeskip \Varid{hs}\codeskip \Varid{cs}){}\<[E]%
\\
\>[3]{}\hsindent{2}{}\<[5]%
\>[5]{}\Rightarrow \Conid{Circuit}\codeskip {}\<[17]%
\>[17]{}\Varid{fs}\codeskip {}\<[21]%
\>[21]{}\Varid{as}\codeskip {}\<[25]%
\>[25]{}\Varid{gs}\codeskip {}\<[29]%
\>[29]{}\Varid{bs}\codeskip {}\<[33]%
\>[33]{}\Varid{nfs}\to \Conid{Circuit}\codeskip {}\<[49]%
\>[49]{}\Varid{gs}\codeskip {}\<[53]%
\>[53]{}\Varid{bs}\codeskip {}\<[57]%
\>[57]{}\Varid{hs}\codeskip {}\<[61]%
\>[61]{}\Varid{cs}\codeskip {}\<[65]%
\>[65]{}\Varid{ngs}\to \Conid{Circuit}\codeskip {}\<[81]%
\>[81]{}\Varid{fs}\codeskip {}\<[85]%
\>[85]{}\Varid{as}\codeskip {}\<[89]%
\>[89]{}\Varid{hs}\codeskip {}\<[93]%
\>[93]{}\Varid{cs}\codeskip {}\<[97]%
\>[97]{}\Varid{nfs}{}\<[E]%
\ColumnHook
\end{hscode}\resethooks
It employs a similar logic to function composition \ensuremath{(\hsdot{\cdot }{.})\mathbin{::}(\Varid{a}\to \Varid{b})\to (\Varid{b}\to \Varid{c})\to (\Varid{a}\to \Varid{c})}.
The resulting type from this constructor uses the input types from the first argument \ensuremath{\Varid{fs}\codeskip \Varid{as}},
and the output types from the second argument \ensuremath{\Varid{hs}\codeskip \Varid{cs}}.
It then forces the constraint that the output type of the first argument and the input type of the second are the same --- \ensuremath{\Varid{gs}\codeskip \Varid{bs}}.

\paragraph{The `Beside' Constructor},
denoted by \ensuremath{\mathbin{<>}} is used to run two circuits at the same time.
The resulting \ensuremath{\Conid{Circuit}} has the types of the two circuits appended together.
\begin{hscode}\SaveRestoreHook
\column{B}{@{}>{\hspre}l<{\hspost}@{}}%
\column{3}{@{}>{\hspre}l<{\hspost}@{}}%
\column{5}{@{}>{\hspre}c<{\hspost}@{}}%
\column{5E}{@{}l@{}}%
\column{9}{@{}>{\hspre}l<{\hspost}@{}}%
\column{12}{@{}>{\hspre}c<{\hspost}@{}}%
\column{12E}{@{}l@{}}%
\column{15}{@{}>{\hspre}l<{\hspost}@{}}%
\column{18}{@{}>{\hspre}l<{\hspost}@{}}%
\column{22}{@{}>{\hspre}l<{\hspost}@{}}%
\column{26}{@{}>{\hspre}l<{\hspost}@{}}%
\column{30}{@{}>{\hspre}l<{\hspost}@{}}%
\column{34}{@{}>{\hspre}l<{\hspost}@{}}%
\column{50}{@{}>{\hspre}l<{\hspost}@{}}%
\column{54}{@{}>{\hspre}l<{\hspost}@{}}%
\column{58}{@{}>{\hspre}l<{\hspost}@{}}%
\column{62}{@{}>{\hspre}l<{\hspost}@{}}%
\column{66}{@{}>{\hspre}l<{\hspost}@{}}%
\column{E}{@{}>{\hspre}l<{\hspost}@{}}%
\>[3]{}(\mathbin{<>})\mathbin{::}{}\<[12]%
\>[12]{}({}\<[12E]%
\>[15]{}\Conid{DataStore}'\codeskip \Varid{fs}\codeskip \Varid{as},\Conid{DataStore}'\codeskip \Varid{gs}\codeskip \Varid{bs},\Conid{DataStore}'\codeskip \Varid{hs}\codeskip \Varid{cs},\Conid{DataStore}'\codeskip \Varid{is}\codeskip \Varid{ds}){}\<[E]%
\\
\>[3]{}\hsindent{2}{}\<[5]%
\>[5]{}\Rightarrow {}\<[5E]%
\>[9]{}\Conid{Circuit}\codeskip {}\<[18]%
\>[18]{}\Varid{fs}\codeskip {}\<[22]%
\>[22]{}\Varid{as}\codeskip {}\<[26]%
\>[26]{}\Varid{gs}\codeskip {}\<[30]%
\>[30]{}\Varid{bs}\codeskip {}\<[34]%
\>[34]{}\Varid{nfs}\to \Conid{Circuit}\codeskip {}\<[50]%
\>[50]{}\Varid{hs}\codeskip {}\<[54]%
\>[54]{}\Varid{cs}\codeskip {}\<[58]%
\>[58]{}\Varid{is}\codeskip {}\<[62]%
\>[62]{}\Varid{ds}\codeskip {}\<[66]%
\>[66]{}\Varid{nhs}{}\<[E]%
\\
\>[3]{}\hsindent{2}{}\<[5]%
\>[5]{}\to {}\<[5E]%
\>[9]{}\Conid{Circuit}\codeskip {}\<[18]%
\>[18]{}(\Varid{fs}:\!\!+\!\!+\Varid{hs})\codeskip (\Varid{as}:\!\!+\!\!+\Varid{cs})\codeskip (\Varid{gs}:\!\!+\!\!+\Varid{is})\codeskip (\Varid{bs}:\!\!+\!\!+\Varid{ds})\codeskip (\Varid{nfs}:\!\!+\Varid{nhs}){}\<[E]%
\ColumnHook
\end{hscode}\resethooks
This constructor works by making use of the \ensuremath{:\!\!+\!\!+} type family to append the input and output type list of the left constructor to those of the right constructor.
It also makes use of the \ensuremath{:\!\!+} type family to sum the number of inputs.

Tasks are made using a smart constructor \ensuremath{\Varid{task}}, which requires a type level \ensuremath{\Conid{Length}}.
To save boiler-plate, \ensuremath{\Conid{CircuitFlow}} also provides more handy task smart constructors such as \ensuremath{\Varid{functionTask}}.
This particular smart constructor allows a simple \ensuremath{\Varid{a}\to \Varid{b}} function to be promoted to a task.
It comes in useful returning to the music preprocessing example as it simplifies the definition of a task that finds the top ten songs or artists: \ensuremath{\Varid{functionTask}\codeskip (\Varid{take}\codeskip \mathrm{10})}.

\subsection{\ensuremath{\Conid{CircuitFlow}} in Action}

\noindent\begin{minipage}{\linewidth}
\begin{hscode}\SaveRestoreHook
\column{B}{@{}>{\hspre}l<{\hspost}@{}}%
\column{20}{@{}>{\hspre}l<{\hspost}@{}}%
\column{33}{@{}>{\hspre}c<{\hspost}@{}}%
\column{33E}{@{}l@{}}%
\column{38}{@{}>{\hspre}c<{\hspost}@{}}%
\column{38E}{@{}l@{}}%
\column{39}{@{}>{\hspre}c<{\hspost}@{}}%
\column{39E}{@{}l@{}}%
\column{43}{@{}>{\hspre}l<{\hspost}@{}}%
\column{56}{@{}>{\hspre}l<{\hspost}@{}}%
\column{E}{@{}>{\hspre}l<{\hspost}@{}}%
\>[B]{}\Varid{preProcPipeline}\mathrel{=}{}\<[20]%
\>[20]{}\Varid{organiseIns}{}\<[33]%
\>[33]{}<\!\!\!\!-\!\!\!\!>{}\<[33E]%
\>[38]{}({}\<[38E]%
\>[43]{}(\Varid{aggSongs}{}\<[56]%
\>[56]{}<\!\!\!\!-\!\!\!\!>\Varid{top10}\codeskip \text{\ttfamily \char34 t10s.csv\char34}){}\<[E]%
\\
\>[38]{}\hsindent{1}{}\<[39]%
\>[39]{}\mathbin{<>}{}\<[39E]%
\>[43]{}(\Varid{aggArtists}{}\<[56]%
\>[56]{}<\!\!\!\!-\!\!\!\!>\Varid{top10}\codeskip \text{\ttfamily \char34 t10a.csv\char34})){}\<[E]%
\ColumnHook
\end{hscode}\resethooks
\end{minipage}

The above \ensuremath{\Conid{CircuitFlow}} circuit solves the music processing example.
\ensuremath{\Varid{organiseIns}} replicates the input values so that they are passed into both \ensuremath{\Varid{aggSongs}} and \ensuremath{\Varid{aggArtists}}.
Again, it can be seen how this structure of tasks directly correlates with the dataflow diagram previously seen in Figure~\ref{fig:example-song-pre-proc}.
This helps to make it easier when designing circuits as it can be constructed visually level by level.

\subsection{mapC operator}
Currently a circuit has a static design: once created it cannot change.
There are times when this could be a flaw in the language.
For example, when there is a dynamic number of inputs.
\ensuremath{\Conid{CircuitFlow}}'s \ensuremath{\Varid{mapC}} allows for dynamic circuits.
This constructor maps a circuit on an input containing a list of items.
The input is fed one at a time into the inner circuit, accumulated back into a list, and then output.

\begin{hscode}\SaveRestoreHook
\column{B}{@{}>{\hspre}l<{\hspost}@{}}%
\column{3}{@{}>{\hspre}l<{\hspost}@{}}%
\column{E}{@{}>{\hspre}l<{\hspost}@{}}%
\>[B]{}\Varid{mapC}\mathbin{::}(\Conid{DataStore}'\codeskip '[\mskip1.5mu \Varid{f}\mskip1.5mu]\codeskip '[\mskip1.5mu [\mskip1.5mu \Varid{a}\mskip1.5mu]\mskip1.5mu],\Conid{DataStore}\codeskip \Varid{g}\codeskip [\mskip1.5mu \Varid{b}\mskip1.5mu]){}\<[E]%
\\
\>[B]{}\hsindent{3}{}\<[3]%
\>[3]{}\Rightarrow \Conid{Circuit}\codeskip '[\mskip1.5mu \Conid{Var}\mskip1.5mu]\codeskip '[\mskip1.5mu \Varid{a}\mskip1.5mu]\codeskip '[\mskip1.5mu \Conid{Var}\mskip1.5mu]\codeskip '[\mskip1.5mu \Varid{b}\mskip1.5mu]\codeskip \Conid{N1}\to \Conid{Circuit}\codeskip '[\mskip1.5mu \Varid{f}\mskip1.5mu]\codeskip '[\mskip1.5mu [\mskip1.5mu \Varid{a}\mskip1.5mu]\mskip1.5mu]\codeskip '[\mskip1.5mu \Varid{g}\mskip1.5mu]\codeskip '[\mskip1.5mu [\mskip1.5mu \Varid{b}\mskip1.5mu]\mskip1.5mu]\codeskip \Conid{N1}{}\<[E]%
\ColumnHook
\end{hscode}\resethooks


\section{\ensuremath{\Conid{CircuitFlow}} Under the Hood}\label{sec:backend}

This section explores the embedding of the \ensuremath{\Conid{CircuitFlow}} language into Haskell and how it is translated down to be executed.

\subsection{Circuit API}
The constructors for the language are actually \textit{smart constructors} \cite{SVENNINGSSON2015143}, providing a more elegant way to build the \acs{AST} that represents the circuit. They bring the benefits of extensibility and modularity usually found in a shallow embedding, while still having a fixed core \acs{AST} that can be used for interpretation.

\paragraph{IFunctor}
The fixed core \acs{AST} is implemented via a jacked up version of the traditional capturing of an abstract datatype as a fixed \ensuremath{\Conid{Functor}} story~\cite{foldingDSLs}. Instead of \ensuremath{\Conid{Functor}}, a type class called \ensuremath{\Conid{IFunctor}}~\cite{mcbride2011functional} (also known as \ensuremath{\Conid{HFunctor}}~\cite{10.1145/1328438.1328475}) is used as it is able to maintain the type indices, which in the case of \ensuremath{\Conid{CircuitFlow}}, are the all important dependency phantom type parameters. \ensuremath{\Conid{IFunctor}} can be thought of as a \ensuremath{\Conid{Functor}} transformer: it is able to change the structure of a \ensuremath{\Conid{Functor}}, whilst preserving the values inside it.
\ensuremath{\Conid{IFunctor}}s can also be used to mark recursive points of data types, as long as they are paired with a matching \ensuremath{\Conid{IFix}} to tie the recursive knot.
As \ensuremath{\Conid{Circuit}} has five type parameters, it needs \ensuremath{\Conid{IFunctor}_{5}} and \ensuremath{\Conid{IFix}_{5}}.


\begin{multicols}{2}
\begin{hscode}\SaveRestoreHook
\column{B}{@{}>{\hspre}l<{\hspost}@{}}%
\column{3}{@{}>{\hspre}l<{\hspost}@{}}%
\column{5}{@{}>{\hspre}l<{\hspost}@{}}%
\column{E}{@{}>{\hspre}l<{\hspost}@{}}%
\>[3]{}\mathbf{type}\codeskip (\leadsto )\codeskip \Varid{f}\codeskip \Varid{g}\mathrel{=}\forall \Varid{a}\hsforall \hsdot{\cdot }{.}\Varid{f}\codeskip \Varid{a}\to \Varid{g}\codeskip \Varid{a}{}\<[E]%
\\[\blanklineskip]%
\>[3]{}\mathbf{class}\codeskip \Conid{IFunctor}\codeskip \Varid{iF}\codeskip \mathbf{where}{}\<[E]%
\\
\>[3]{}\hsindent{2}{}\<[5]%
\>[5]{}\Varid{imap}\mathbin{::}(\Varid{f}\leadsto \Varid{g})\to \Varid{iF}\codeskip \Varid{f}\leadsto \Varid{iF}\codeskip \Varid{g}{}\<[E]%
\\[\blanklineskip]%
\>[3]{}\mathbf{newtype}\codeskip \Conid{IFix}\codeskip \Varid{iF}\codeskip \Varid{a}{}\<[E]%
\\
\>[3]{}\hsindent{2}{}\<[5]%
\>[5]{}\mathrel{=}\Conid{IIn}\codeskip (\Varid{iF}\codeskip (\Conid{IFix}\codeskip \Varid{iF})\codeskip \Varid{a}){}\<[E]%
\ColumnHook
\end{hscode}\resethooks
\begin{hscode}\SaveRestoreHook
\column{B}{@{}>{\hspre}l<{\hspost}@{}}%
\column{3}{@{}>{\hspre}l<{\hspost}@{}}%
\column{5}{@{}>{\hspre}l<{\hspost}@{}}%
\column{7}{@{}>{\hspre}l<{\hspost}@{}}%
\column{E}{@{}>{\hspre}l<{\hspost}@{}}%
\>[3]{}\mathbf{class}\codeskip \Conid{IFunctor}_{5}\codeskip \Varid{iF}\codeskip \mathbf{where}{}\<[E]%
\\
\>[3]{}\hsindent{2}{}\<[5]%
\>[5]{}\Varid{imap}_{5}{}\<[E]%
\\
\>[5]{}\hsindent{2}{}\<[7]%
\>[7]{}\mathbin{::}(\forall \Varid{a}\hsforall \mathbin{...}\Varid{e}\hsdot{\cdot }{.}\Varid{f}\codeskip \Varid{a}\mathbin{...}\Varid{e}\to \Varid{g}\codeskip \Varid{a}\mathbin{...}\Varid{e}){}\<[E]%
\\
\>[5]{}\hsindent{2}{}\<[7]%
\>[7]{}\to \Varid{iF}\codeskip \Varid{f}\codeskip \Varid{a}\mathbin{...}\Varid{e}\to \Varid{iF}\codeskip \Varid{g}\codeskip \Varid{a}\mathbin{...}\Varid{e}{}\<[E]%
\\
\>[3]{}\mathbf{newtype}\codeskip \Conid{IFix}_{5}\codeskip \Varid{iF}\codeskip \Varid{a}\mathbin{...}\Varid{e}{}\<[E]%
\\
\>[3]{}\hsindent{2}{}\<[5]%
\>[5]{}\mathrel{=}\Conid{IIn}_{5}\codeskip (\Varid{iF}\codeskip (\Conid{IFix}_{5}\codeskip \Varid{iF})\codeskip \Varid{a}\mathbin{...}\Varid{e}){}\<[E]%
\ColumnHook
\end{hscode}\resethooks
\end{multicols}

\paragraph{Indexed Data types \`{a} la carte}
When building an \ac{eDSL} one problem that becomes quickly prevalent is the so called \textit{Expression Problem}~\cite{wadler_1998}. A popular solution is \textit{Data types \`{a} la carte}~\cite{swierstra_2008}: it combines constructors using the co-product of their signatures.
This technique makes use of standard functors, however, an approach using \ensuremath{\Conid{IFunctor}}s is described in \textit{Compositional data types}~\cite{10.1145/2036918.2036930}. This approach is upgraded further to add support for five type indices:
\begin{hscode}\SaveRestoreHook
\column{B}{@{}>{\hspre}l<{\hspost}@{}}%
\column{3}{@{}>{\hspre}l<{\hspost}@{}}%
\column{5}{@{}>{\hspre}c<{\hspost}@{}}%
\column{5E}{@{}l@{}}%
\column{9}{@{}>{\hspre}l<{\hspost}@{}}%
\column{18}{@{}>{\hspre}l<{\hspost}@{}}%
\column{21}{@{}>{\hspre}l<{\hspost}@{}}%
\column{59}{@{}>{\hspre}l<{\hspost}@{}}%
\column{68}{@{}>{\hspre}l<{\hspost}@{}}%
\column{E}{@{}>{\hspre}l<{\hspost}@{}}%
\>[3]{}\mathbf{data}\codeskip (\Varid{iF}:\!\!+\!\!:\Varid{iG})\codeskip {}\<[21]%
\>[21]{}(\Varid{f'}\mathbin{::}\Varid{i}\to \Varid{j}\to \Varid{k}\to \Varid{l}\to \Varid{m}\to \mathbin{*})\codeskip (\Varid{a}\mathbin{::}\Varid{i})\mathbin{...}(\Varid{e}\mathbin{::}\Varid{m}){}\<[E]%
\\
\>[3]{}\hsindent{2}{}\<[5]%
\>[5]{}\mathrel{=}{}\<[5E]%
\>[9]{}\Conid{L}\mathbin{::}\Varid{iF}\codeskip {}\<[18]%
\>[18]{}\Varid{f'}\codeskip \Varid{a}\mathbin{...}\Varid{e}\to (\Varid{iF}:\!\!+\!\!:\Varid{iG})\codeskip \Varid{f'}\codeskip \Varid{a}\mathbin{...}\Varid{e}\mid {}\<[59]%
\>[59]{}\Conid{R}\mathbin{::}\Varid{iG}\codeskip {}\<[68]%
\>[68]{}\Varid{f'}\codeskip \Varid{a}\mathbin{...}\Varid{e}\to (\Varid{iF}:\!\!+\!\!:\Varid{iG})\codeskip \Varid{f'}\codeskip \Varid{a}\mathbin{...}\Varid{e}{}\<[E]%
\ColumnHook
\end{hscode}\resethooks
Using the \ensuremath{:\!\!+\!\!:} operator comes with problem of many \ensuremath{\Conid{L}}'s and \ensuremath{\Conid{R}}'s, when creating the \acs{AST}. The solution, extended from \cite{swierstra_2008} to also accommodate five type parameters, is to introduce a type class \ensuremath{:\prec:} that injects them automatically.

Data types for each constructor can now be defined individually. The \ensuremath{\Conid{Then}} (\ensuremath{<\!\!\!\!-\!\!\!\!>}) constructor is used as an example, however, the process can be applied to all constructors in the language.
\begin{hscode}\SaveRestoreHook
\column{B}{@{}>{\hspre}l<{\hspost}@{}}%
\column{3}{@{}>{\hspre}l<{\hspost}@{}}%
\column{5}{@{}>{\hspre}l<{\hspost}@{}}%
\column{7}{@{}>{\hspre}c<{\hspost}@{}}%
\column{7E}{@{}l@{}}%
\column{11}{@{}>{\hspre}l<{\hspost}@{}}%
\column{14}{@{}>{\hspre}l<{\hspost}@{}}%
\column{19}{@{}>{\hspre}c<{\hspost}@{}}%
\column{19E}{@{}l@{}}%
\column{22}{@{}>{\hspre}l<{\hspost}@{}}%
\column{23}{@{}>{\hspre}l<{\hspost}@{}}%
\column{43}{@{}>{\hspre}l<{\hspost}@{}}%
\column{56}{@{}>{\hspre}l<{\hspost}@{}}%
\column{63}{@{}>{\hspre}l<{\hspost}@{}}%
\column{E}{@{}>{\hspre}l<{\hspost}@{}}%
\>[3]{}\mathbf{data}\codeskip \Conid{Then}\codeskip {}\<[14]%
\>[14]{}(\Varid{iF}{}\<[19]%
\>[19]{}\mathbin{::}{}\<[19E]%
\>[23]{}[\mskip1.5mu \mathbin{*}\to \mathbin{*}\mskip1.5mu]\to [\mskip1.5mu \mathbin{*}\mskip1.5mu]\to {}\<[43]%
\>[43]{}[\mskip1.5mu \mathbin{*}\to \mathbin{*}\mskip1.5mu]\to [\mskip1.5mu \mathbin{*}\mskip1.5mu]\to {}\<[63]%
\>[63]{}\Conid{Nat}\to \mathbin{*})\codeskip {}\<[E]%
\\
\>[14]{}(\Varid{insS}{}\<[22]%
\>[22]{}\mathbin{::}[\mskip1.5mu \mathbin{*}\to \mathbin{*}\mskip1.5mu])\codeskip (\Varid{insT}{}\<[43]%
\>[43]{}\mathbin{::}[\mskip1.5mu \mathbin{*}\mskip1.5mu])\codeskip {}\<[E]%
\\
\>[14]{}(\Varid{outsS}{}\<[22]%
\>[22]{}\mathbin{::}[\mskip1.5mu \mathbin{*}\to \mathbin{*}\mskip1.5mu])\codeskip (\Varid{outsT}{}\<[43]%
\>[43]{}\mathbin{::}[\mskip1.5mu \mathbin{*}\mskip1.5mu])\codeskip (\Varid{nins}\mathbin{::}\Conid{Nat})\codeskip \mathbf{where}{}\<[E]%
\\
\>[3]{}\hsindent{2}{}\<[5]%
\>[5]{}\Conid{Then}\mathbin{::}(\Conid{DataStore}'\codeskip \Varid{fs}\codeskip \Varid{as},\Conid{DataStore}'\codeskip \Varid{gs}\codeskip \Varid{bs},\Conid{DataStore}'\codeskip \Varid{hs}\codeskip \Varid{cs}){}\<[E]%
\\
\>[5]{}\hsindent{2}{}\<[7]%
\>[7]{}\Rightarrow {}\<[7E]%
\>[11]{}\Varid{iF}\codeskip \Varid{fs}\codeskip \Varid{as}\codeskip \Varid{gs}\codeskip \Varid{bs}\codeskip \Varid{nfs}\to \Varid{iF}\codeskip \Varid{gs}\codeskip \Varid{bs}\codeskip \Varid{hs}\codeskip \Varid{cs}\codeskip \Varid{ngs}\to {}\<[56]%
\>[56]{}\Conid{Then}\codeskip \Varid{iF}\codeskip \Varid{fs}\codeskip \Varid{as}\codeskip \Varid{hs}\codeskip \Varid{cs}\codeskip \Varid{nfs}{}\<[E]%
\ColumnHook
\end{hscode}\resethooks
Each \ensuremath{\Varid{iF}} denotes the recursive points in the data type, with the subsequent type arguments mirroring those seen in Section~\ref{sec:lang-circuit-constructors}.
A corresponding \ensuremath{\Conid{IFunctor}_{5}} instance formalises the points of recursion, by describing how to transform the structure inside it.
The smart constructor, that injects the \ensuremath{\Conid{L}}'s and \ensuremath{\Conid{R}}'s automatically can be defined for \ensuremath{\Conid{Then}} adding one extra constraint, to the constructor defined in Section~\ref{sec:lang-circuit-constructors} (\ensuremath{\Conid{Then}:\prec:\Varid{iF}}), allowing the smart constructor to produce a node in the \acs{AST} for any sum of data types, that includes the \ensuremath{\Conid{Then}} data type.

\paragraph{Representing a Circuit}
Once each constructor has been defined, they can be combined together to form the \ensuremath{\Conid{CircuitF}} type to represent a circuit. \ensuremath{\Conid{IFix}_{5}} then ties the recursive knot to define the \ensuremath{\Conid{Circuit}} type.
\begin{hscode}\SaveRestoreHook
\column{B}{@{}>{\hspre}l<{\hspost}@{}}%
\column{3}{@{}>{\hspre}l<{\hspost}@{}}%
\column{E}{@{}>{\hspre}l<{\hspost}@{}}%
\>[3]{}\mathbf{type}\codeskip \Conid{CircuitF}\mathrel{=}\Conid{Id}:\!\!+\!\!:\Conid{Replicate}:\!\!+\!\!:\Conid{Then}:\!\!+\!\!:\mathbin{...}:\!\!+\!\!:\Conid{Task}:\!\!+\!\!:\Conid{Map}{}\<[E]%
\\
\>[3]{}\mathbf{type}\codeskip \Conid{Circuit}\mathrel{=}\Conid{IFix}_{5}\codeskip \Conid{CircuitF}{}\<[E]%
\ColumnHook
\end{hscode}\resethooks
Now that it is possible to build a \ensuremath{\Conid{Circuit}}, which can be considered a specification for how to execute a set of tasks, there needs to be a mechanism in place to execute the specification.

\subsection{Network Typeclass}

A \ensuremath{\Conid{Network}} represents a mechanism for executing the computation described by a \ensuremath{\Conid{Circuit}}.
To allow for multiple execution mechanisms, a \ensuremath{\Conid{Network}} type class defines the key features each network requires:
\begin{hscode}\SaveRestoreHook
\column{B}{@{}>{\hspre}l<{\hspost}@{}}%
\column{3}{@{}>{\hspre}l<{\hspost}@{}}%
\column{5}{@{}>{\hspre}l<{\hspost}@{}}%
\column{19}{@{}>{\hspre}c<{\hspost}@{}}%
\column{19E}{@{}l@{}}%
\column{23}{@{}>{\hspre}l<{\hspost}@{}}%
\column{26}{@{}>{\hspre}l<{\hspost}@{}}%
\column{28}{@{}>{\hspre}l<{\hspost}@{}}%
\column{32}{@{}>{\hspre}l<{\hspost}@{}}%
\column{49}{@{}>{\hspre}l<{\hspost}@{}}%
\column{53}{@{}>{\hspre}l<{\hspost}@{}}%
\column{80}{@{}>{\hspre}l<{\hspost}@{}}%
\column{E}{@{}>{\hspre}l<{\hspost}@{}}%
\>[3]{}\mathbf{class}\codeskip \Conid{Network}\codeskip \Varid{n}\codeskip \mathbf{where}{}\<[E]%
\\
\>[3]{}\hsindent{2}{}\<[5]%
\>[5]{}\Varid{startNetwork}{}\<[19]%
\>[19]{}\mathbin{::}{}\<[19E]%
\>[23]{}\Conid{Circuit}\codeskip {}\<[32]%
\>[32]{}\Varid{insS}\codeskip \Varid{insT}\codeskip \Varid{outsS}\codeskip \Varid{outsT}\codeskip \Varid{nIns}{}\<[E]%
\\
\>[19]{}\to {}\<[19E]%
\>[23]{}\Conid{IO}\codeskip {}\<[28]%
\>[28]{}(\Varid{n}\codeskip {}\<[32]%
\>[32]{}\Varid{insS}\codeskip \Varid{insT}\codeskip \Varid{outsS}\codeskip \Varid{outsT}){}\<[E]%
\\
\>[3]{}\hsindent{2}{}\<[5]%
\>[5]{}\Varid{stopNetwork}{}\<[19]%
\>[19]{}\mathbin{::}{}\<[19E]%
\>[23]{}\Varid{n}\codeskip {}\<[26]%
\>[26]{}\Varid{insS}\codeskip \Varid{insT}\codeskip \Varid{outsS}\codeskip \Varid{outsT}{}\<[49]%
\>[49]{}\to {}\<[53]%
\>[53]{}\Conid{IO}\codeskip (){}\<[E]%
\\
\>[3]{}\hsindent{2}{}\<[5]%
\>[5]{}\Varid{write}{}\<[19]%
\>[19]{}\mathbin{::}{}\<[19E]%
\>[23]{}\Conid{IHList}\codeskip \Varid{insS}\codeskip \Varid{insT}{}\<[49]%
\>[49]{}\to \Varid{n}\codeskip \Varid{insS}\codeskip \Varid{insT}\codeskip \Varid{outsS}\codeskip \Varid{outsT}\to {}\<[80]%
\>[80]{}\Conid{IO}\codeskip (){}\<[E]%
\\
\>[3]{}\hsindent{2}{}\<[5]%
\>[5]{}\Varid{read}{}\<[19]%
\>[19]{}\mathbin{::}{}\<[19E]%
\>[23]{}\Varid{n}\codeskip \Varid{insS}\codeskip \Varid{insT}\codeskip \Varid{outsS}\codeskip \Varid{outsT}{}\<[49]%
\>[49]{}\to \Conid{IO}\codeskip (\Conid{IHList}\codeskip \Varid{outsS}\codeskip \Varid{outsT}){}\<[E]%
\ColumnHook
\end{hscode}\resethooks
This type class requires that a network has 4 different functions:
\ensuremath{\Varid{startNetwork}} is responsible for converting the circuit into the underlying representation for a process network: it will be discussed in more detail in Section~\ref{sec:circuit-translation};
\ensuremath{\Varid{stopNetwork}} is for cleaning up the network after it is no longer needed. For example, stopping any threads running. This could be particularly important if embedding a circuit into a larger program, where unused threads could be left hanging;
\ensuremath{\Varid{write}} should take some input values and add them into the network, so that they can be processed;
\ensuremath{\Varid{read}} should retrieve some output values from the network.
\ensuremath{\Varid{nIns}} is required for the translation of \ensuremath{\Conid{Circuit}} to \ensuremath{\Conid{Network}}, therefore it is not inluded in the type of a network.

\subsection{The Basic Network Representation}
A \ensuremath{\Conid{BasicNetwork}} is an implementation of a \ensuremath{\Conid{Network}} that uses a \acf{KPN}.
This means that each task in a circuit will run on its own separate thread, with inputs being passed between them on unbounded channels (from \ensuremath{\Conid{\Conid{Control}.Concurrent}}).
A \ensuremath{\Conid{BasicNetwork}} stores the multiple input and output channels, to do so it leverages a special case of \ensuremath{\Conid{IHList}}.
\begin{hscode}\SaveRestoreHook
\column{B}{@{}>{\hspre}l<{\hspost}@{}}%
\column{3}{@{}>{\hspre}l<{\hspost}@{}}%
\column{5}{@{}>{\hspre}l<{\hspost}@{}}%
\column{15}{@{}>{\hspre}l<{\hspost}@{}}%
\column{18}{@{}>{\hspre}l<{\hspost}@{}}%
\column{E}{@{}>{\hspre}l<{\hspost}@{}}%
\>[3]{}\mathbf{data}\codeskip \Conid{PipeList}\codeskip {}\<[18]%
\>[18]{}(\Varid{fs}\mathbin{::}[\mskip1.5mu \mathbin{*}\to \mathbin{*}\mskip1.5mu])\codeskip (\Varid{as}\mathbin{::}[\mskip1.5mu \mathbin{*}\mskip1.5mu])\codeskip \mathbf{where}{}\<[E]%
\\
\>[3]{}\hsindent{2}{}\<[5]%
\>[5]{}\Conid{PipeCons}{}\<[15]%
\>[15]{}\mathbin{::}\Conid{Chan}\codeskip (\Varid{f}\codeskip \Varid{a})\to \Conid{PipeList}\codeskip \Varid{fs}\codeskip \Varid{as}\to \Conid{PipeList}\codeskip (\Varid{f}\codeskip \,'\!\!\mathbin{:}\codeskip \Varid{fs})\codeskip (\Varid{a}\codeskip \,'\!\!\mathbin{:}\codeskip \Varid{as}){}\<[E]%
\\
\>[3]{}\hsindent{2}{}\<[5]%
\>[5]{}\Conid{PipeNil}{}\<[15]%
\>[15]{}\mathbin{::}\Conid{PipeList}\codeskip '[\mskip1.5mu \mskip1.5mu]\codeskip '[\mskip1.5mu \mskip1.5mu]{}\<[E]%
\ColumnHook
\end{hscode}\resethooks
Using these \ensuremath{\Conid{PipeList}}s, \ensuremath{\Conid{BasicNetwork}} is defined using record syntax allowing for named fields, with accessors automatically generated.
\begin{hscode}\SaveRestoreHook
\column{B}{@{}>{\hspre}l<{\hspost}@{}}%
\column{3}{@{}>{\hspre}l<{\hspost}@{}}%
\column{5}{@{}>{\hspre}l<{\hspost}@{}}%
\column{7}{@{}>{\hspre}l<{\hspost}@{}}%
\column{16}{@{}>{\hspre}l<{\hspost}@{}}%
\column{22}{@{}>{\hspre}l<{\hspost}@{}}%
\column{24}{@{}>{\hspre}l<{\hspost}@{}}%
\column{29}{@{}>{\hspre}l<{\hspost}@{}}%
\column{30}{@{}>{\hspre}l<{\hspost}@{}}%
\column{34}{@{}>{\hspre}l<{\hspost}@{}}%
\column{36}{@{}>{\hspre}l<{\hspost}@{}}%
\column{44}{@{}>{\hspre}l<{\hspost}@{}}%
\column{47}{@{}>{\hspre}l<{\hspost}@{}}%
\column{52}{@{}>{\hspre}l<{\hspost}@{}}%
\column{E}{@{}>{\hspre}l<{\hspost}@{}}%
\>[3]{}\mathbf{data}\codeskip \Conid{BasicNetwork}\codeskip {}\<[22]%
\>[22]{}(\Varid{insS}{}\<[30]%
\>[30]{}\mathbin{::}[\mskip1.5mu \mathbin{*}\to \mathbin{*}\mskip1.5mu])\codeskip {}\<[44]%
\>[44]{}(\Varid{insT}{}\<[52]%
\>[52]{}\mathbin{::}[\mskip1.5mu \mathbin{*}\mskip1.5mu])\codeskip {}\<[E]%
\\
\>[22]{}(\Varid{outsS}{}\<[30]%
\>[30]{}\mathbin{::}[\mskip1.5mu \mathbin{*}\to \mathbin{*}\mskip1.5mu])\codeskip {}\<[44]%
\>[44]{}(\Varid{outsT}{}\<[52]%
\>[52]{}\mathbin{::}[\mskip1.5mu \mathbin{*}\mskip1.5mu])\codeskip \mathbf{where}{}\<[E]%
\\
\>[3]{}\hsindent{2}{}\<[5]%
\>[5]{}\Conid{BasicNetwork}\mathbin{::}\{\mskip1.5mu {}\<[E]%
\\
\>[5]{}\hsindent{2}{}\<[7]%
\>[7]{}\Varid{threads}{}\<[16]%
\>[16]{}\mathbin{::}\Conid{Map}\codeskip {}\<[24]%
\>[24]{}\Conid{TaskUUID}\codeskip {}\<[34]%
\>[34]{}\Conid{ThreadId},{}\<[47]%
\>[47]{}\mbox{\onelinecomment  allows threads to be managed}{}\<[E]%
\\
\>[5]{}\hsindent{2}{}\<[7]%
\>[7]{}\Varid{jobs}{}\<[16]%
\>[16]{}\mathbin{::}\Conid{Map}\codeskip {}\<[24]%
\>[24]{}\Conid{JobUUID}\codeskip {}\<[34]%
\>[34]{}\Conid{JobStatus},{}\<[47]%
\>[47]{}\mbox{\onelinecomment  avoids duplicate job UUIDs}{}\<[E]%
\\
\>[5]{}\hsindent{2}{}\<[7]%
\>[7]{}\Varid{ins}{}\<[16]%
\>[16]{}\mathbin{::}\Conid{PipeList}\codeskip {}\<[29]%
\>[29]{}\Varid{inpS}\codeskip {}\<[36]%
\>[36]{}\Varid{inpT},{}\<[47]%
\>[47]{}\mbox{\onelinecomment  to feed in inputs}{}\<[E]%
\\
\>[5]{}\hsindent{2}{}\<[7]%
\>[7]{}\Varid{outs}{}\<[16]%
\>[16]{}\mathbin{::}\Conid{PipeList}\codeskip {}\<[29]%
\>[29]{}\Varid{outsS}\codeskip {}\<[36]%
\>[36]{}\Varid{outsT}{}\<[47]%
\>[47]{}\mbox{\onelinecomment  to retrieve outputs}{}\<[E]%
\\
\>[3]{}\hsindent{2}{}\<[5]%
\>[5]{}\mskip1.5mu\}\to \Conid{BasicNetwork}\codeskip \Varid{inS}\codeskip \Varid{insT}\codeskip \Varid{outsS}\codeskip \Varid{outsT}{}\<[E]%
\ColumnHook
\end{hscode}\resethooks
The \ensuremath{\Conid{Network}} type instance for a \ensuremath{\Conid{BasicNetwork}} is relatively trivial to implement using \ensuremath{\Conid{\Conid{Control}.Monad}}'s \ensuremath{\keyw{forM\_}} if given a function to transform a \ensuremath{\Conid{Circuit}} to it.
\begin{hscode}\SaveRestoreHook
\column{B}{@{}>{\hspre}l<{\hspost}@{}}%
\column{3}{@{}>{\hspre}l<{\hspost}@{}}%
\column{5}{@{}>{\hspre}l<{\hspost}@{}}%
\column{11}{@{}>{\hspre}l<{\hspost}@{}}%
\column{20}{@{}>{\hspre}l<{\hspost}@{}}%
\column{23}{@{}>{\hspre}l<{\hspost}@{}}%
\column{E}{@{}>{\hspre}l<{\hspost}@{}}%
\>[3]{}\mathbf{instance}\codeskip \Conid{Network}\codeskip \Conid{BasicNetwork}\codeskip \mathbf{where}{}\<[E]%
\\
\>[3]{}\hsindent{2}{}\<[5]%
\>[5]{}\Varid{startNetwork}{}\<[23]%
\>[23]{}\mathrel{=}\Varid{buildBasicNetwork}\mbox{\onelinecomment  Defined soon...}{}\<[E]%
\\
\>[3]{}\hsindent{2}{}\<[5]%
\>[5]{}\Varid{stopNetwork}\codeskip {}\<[20]%
\>[20]{}\Varid{n}{}\<[23]%
\>[23]{}\mathrel{=}\keyw{forM\_}\codeskip (\Varid{threads}\codeskip \Varid{n})\codeskip \Varid{killThread}{}\<[E]%
\\
\>[3]{}\hsindent{2}{}\<[5]%
\>[5]{}\Varid{write}\codeskip \Varid{uuid}\codeskip \Varid{xs}\codeskip {}\<[20]%
\>[20]{}\Varid{n}{}\<[23]%
\>[23]{}\mathrel{=}\Varid{writePipes}\codeskip \Varid{xs}\codeskip (\Varid{ins}\codeskip \Varid{n}){}\<[E]%
\\
\>[3]{}\hsindent{2}{}\<[5]%
\>[5]{}\Varid{read}\codeskip {}\<[11]%
\>[11]{}\Varid{n}{}\<[23]%
\>[23]{}\mathrel{=}\Varid{readPipes}\codeskip (\Varid{outs}\codeskip \Varid{n}){}\<[E]%
\ColumnHook
\end{hscode}\resethooks
The \ensuremath{\Varid{writePipes}} function will input a list of values into each of the respective pipes.
The \ensuremath{\Varid{readPipes}} function will make a blocking call to each channel to read an output from it.
This function will block till an output is read from every output channel.

\subsection{Translation to a BasicNetwork}\label{sec:circuit-translation}

There is now a representation for a \ensuremath{\Conid{Circuit}} that the user will build, and a representation used to execute the \ensuremath{\Conid{Circuit}}.
However, there is no mechanism to convert between them. This can be achieved by folding the circuit data type into a network.
This fold, however, will need to create threads and channels, both of which are \ensuremath{\Conid{IO}} actions, and of course it will also need to deal with the numerous type parameters of \ensuremath{\Conid{Circuit}}. Such requirements lead to an exciting take on the \textit{catamorphism} method for performing generalised folding of an abstract datatype.

\subsubsection{Indexed Monadic Catamorphism}
The use of a catamorphism removes the recursion from any folding of the datatype.
This means that the algebra can focus on one layer at a time.
This also ensures that there is no re-computation of recursive calls, as this is all handled by the catamorphism.
\ensuremath{\Varid{icata}} is able to fold an \ensuremath{\Conid{IFix}\codeskip \Varid{iF}\codeskip \Varid{a}} and produce an item of type \ensuremath{\Varid{f}\codeskip \Varid{a}}.
It uses the algebra argument as a specification of how to transform a single layer of the datatype.
Normal catamorphisms can use monadic computations if defined as follows:
\begin{hscode}\SaveRestoreHook
\column{B}{@{}>{\hspre}l<{\hspost}@{}}%
\column{3}{@{}>{\hspre}l<{\hspost}@{}}%
\column{10}{@{}>{\hspre}l<{\hspost}@{}}%
\column{E}{@{}>{\hspre}l<{\hspost}@{}}%
\>[3]{}\Varid{cataM}{}\<[10]%
\>[10]{}\mathbin{::}(\Conid{Traversable}\codeskip \Varid{f},\Conid{Monad}\codeskip \Varid{m})\Rightarrow (\forall \Varid{a}\hsforall \hsdot{\cdot }{.}\Varid{f}\codeskip \Varid{a}\to \Varid{m}\codeskip \Varid{a})\to \Conid{Fix}\codeskip \Varid{f}\to \Varid{m}\codeskip \Varid{a}{}\<[E]%
\\
\>[3]{}\Varid{cataM}\codeskip \Varid{algM}\codeskip (\Conid{In}\codeskip \Varid{x})\mathrel{=}\Varid{algM}\rbind \Varid{mapM}\codeskip (\Varid{cataM}\codeskip \Varid{algM})\codeskip \Varid{x}{}\<[E]%
\ColumnHook
\end{hscode}\resethooks
This monadic catamorphism~\cite{monadic_cata} follows a similar pattern to a standard catamorphism, but instead uses functions such as a monadic map --- \ensuremath{\Varid{mapM}\mathbin{::}\Conid{Monad}\codeskip \Varid{m}\Rightarrow (\Varid{a}\to \Varid{m}\codeskip \Varid{b})\to \Varid{f}\codeskip \Varid{a}\to \Varid{m}\codeskip (\Varid{f}\codeskip \Varid{b})}.
This allows the monadic catamorphism to be applied recursively on the data type being folded.

A similar technique can also be applied to indexed catamorphisms to gain a monadic version~\cite{10.1145/2036918.2036930}, however, to do so an indexed monadic map has to be introduced. \ensuremath{\Varid{imapM}} is the indexed equivalent of \ensuremath{\Varid{mapM}}, it performs a natural transformation, but is capable of also using monadic computation.
This is included in the \ensuremath{\Conid{IFunctor}} type class, and facilitates the definition of \ensuremath{\Varid{icataM}}.

For \ensuremath{\Conid{Circuit}}, there is one final step that needs to be done: accommodating the five type parameters. To do this, \ensuremath{\Conid{IFunctor}}'s \ensuremath{\Varid{imapM}} gets gifted the type parameters to complete the \ensuremath{\Conid{IFunctor}_{5}} class and allow the definition of \ensuremath{\Varid{icataM}_{5}}.

\subsubsection{BuildNetworkAlg}
The final piece of the translation puzzle is an algebra for the fold. However, a standard algebra will not be able to complete this transformation.
Consider an example \ensuremath{\Conid{Circuit}} with two tasks executed in sequence: \ensuremath{\Varid{task1}<\!\!\!\!-\!\!\!\!>\Varid{task2}}. In a standard algebra, both sides of the \ensuremath{\Conid{Then}} constructor would be evaluated independently.
In this case it would produce two disjoint networks, both with their own input and output channels.
The algebra for \ensuremath{\Conid{Then}}, would then need to join the output channels of \ensuremath{\Varid{task1}} with the input channels of \ensuremath{\Varid{task2}}.
However, it is not possible to join channels together.
Instead, the output channels from \ensuremath{\Varid{task1}} need to be accessible when creating \ensuremath{\Varid{task2}}.
This is referred to as a \textit{context-sensitive} or \textit{accumulating} fold.
An accumulating fold forms series of nested functions, that collapse to give a final value once the base case has been applied.
A simple example of an accumulating fold could be, implementing \ensuremath{\Varid{foldl}} in terms of \ensuremath{\Varid{foldr}}.

To be able to have an accumulating fold inside an indexed catamorphism a carrier data type is required to wrap up this function.
This carrier, which shall be named \ensuremath{\Conid{AccuN}}, contains a function that when given a network that has been accumulated up to that point,
then it is able to produce a network including the next layer in a circuit.
This can be likened to the lambda function given to \ensuremath{\Varid{foldr}}, when defining \ensuremath{\Varid{foldl}}.
The type of the layer being folded will be \ensuremath{\Conid{Circuit}\codeskip \Varid{a}\codeskip \Varid{b}\codeskip \Varid{c}\codeskip \Varid{d}\codeskip \Varid{e}}.
\begin{hscode}\SaveRestoreHook
\column{B}{@{}>{\hspre}l<{\hspost}@{}}%
\column{3}{@{}>{\hspre}l<{\hspost}@{}}%
\column{5}{@{}>{\hspre}l<{\hspost}@{}}%
\column{E}{@{}>{\hspre}l<{\hspost}@{}}%
\>[3]{}\mathbf{newtype}\codeskip \Conid{AccuN}\codeskip \Varid{n}\codeskip \Varid{asS}\codeskip \Varid{asT}\codeskip \Varid{a}\codeskip \Varid{b}\codeskip \Varid{c}\codeskip \Varid{d}\codeskip \Varid{e}\mathrel{=}\Conid{AccuN}{}\<[E]%
\\
\>[3]{}\hsindent{2}{}\<[5]%
\>[5]{}\{\mskip1.5mu \Varid{unAccuN}\mathbin{::}\Varid{n}\codeskip \Varid{asS}\codeskip \Varid{asT}\codeskip \Varid{a}\codeskip \Varid{b}\to \Conid{IO}\codeskip (\Varid{n}\codeskip \Varid{asS}\codeskip \Varid{asT}\codeskip \Varid{c}\codeskip \Varid{d})\mskip1.5mu\}{}\<[E]%
\ColumnHook
\end{hscode}\resethooks
This newtype has two additional type parameters at the beginning, namely: \ensuremath{\Varid{asS}} and \ensuremath{\Varid{asT}}.
They represent the input types to the initial circuit.
Since the accumulating fold will work layer by layer from the top downwards, these types will remain constant and never change throughout the fold.

\paragraph{Classy Algebra}
To ensure that the approach remains modular, the algebra takes the form of a type class: the interpretation of a new constructor is just a new type class instance.
\begin{hscode}\SaveRestoreHook
\column{B}{@{}>{\hspre}l<{\hspost}@{}}%
\column{3}{@{}>{\hspre}l<{\hspost}@{}}%
\column{5}{@{}>{\hspre}l<{\hspost}@{}}%
\column{22}{@{}>{\hspre}c<{\hspost}@{}}%
\column{22E}{@{}l@{}}%
\column{26}{@{}>{\hspre}l<{\hspost}@{}}%
\column{33}{@{}>{\hspre}l<{\hspost}@{}}%
\column{E}{@{}>{\hspre}l<{\hspost}@{}}%
\>[3]{}\mathbf{class}\codeskip (\Conid{Network}\codeskip \Varid{n},\Conid{IFunctor}_{5}\codeskip \Varid{iF})\Rightarrow \Conid{BuildNetworkAlg}\codeskip \Varid{n}\codeskip \Varid{iF}\codeskip \mathbf{where}{}\<[E]%
\\
\>[3]{}\hsindent{2}{}\<[5]%
\>[5]{}\Varid{buildNetworkAlg}{}\<[22]%
\>[22]{}\mathbin{::}{}\<[22E]%
\>[26]{}\Varid{iF}\codeskip ({}\<[33]%
\>[33]{}\Conid{AccuN}\codeskip \Varid{n}\codeskip \Varid{asS}\codeskip \Varid{asT})\codeskip \Varid{bsS}\codeskip \Varid{bsT}\codeskip \Varid{csS}\codeskip \Varid{csT}\codeskip \Varid{nbs}{}\<[E]%
\\
\>[22]{}\to {}\<[22E]%
\>[26]{}\Conid{IO}\codeskip (({}\<[33]%
\>[33]{}\Conid{AccuN}\codeskip \Varid{n}\codeskip \Varid{asS}\codeskip \Varid{asT})\codeskip \Varid{bsS}\codeskip \Varid{bsT}\codeskip \Varid{csS}\codeskip \Varid{csT}\codeskip \Varid{nbs}){}\<[E]%
\ColumnHook
\end{hscode}\resethooks
This algebra type class takes two parameters: \ensuremath{\Varid{n}} and \ensuremath{\Varid{iF}}. The \ensuremath{\Varid{n}} is constrained to have a \ensuremath{\Conid{Network}} instance, this allows the same algebra to be used for defining folds for multiple network types.
The \ensuremath{\Varid{iF}} is the \ensuremath{\Conid{IFunctor}} that this instance is being defined for, an example is \ensuremath{\Conid{Then}} or \ensuremath{\Conid{Id}}.
This algebra uses the \ensuremath{\Conid{AccuN}} data type to perform an accumulating fold.
The input to the algebra is an \ensuremath{\Conid{IFunctor}} with the inner elements containing values of type \ensuremath{\Conid{AccuN}}.
The function can be retrieved from inside \ensuremath{\Conid{AccuN}} to perform steps that are dependent on the previous, for example, in the \ensuremath{\Conid{Then}} constructor.

\paragraph{The Initial Network}
Given the use of an accumulating fold, one important question needs to be answered: what happens on the first layer? The fold needs an \ensuremath{\Varid{initialNetwork}} that has matching input and output types:
\begin{hscode}\SaveRestoreHook
\column{B}{@{}>{\hspre}l<{\hspost}@{}}%
\column{3}{@{}>{\hspre}l<{\hspost}@{}}%
\column{5}{@{}>{\hspre}l<{\hspost}@{}}%
\column{E}{@{}>{\hspre}l<{\hspost}@{}}%
\>[3]{}\Varid{initialNetwork}{}\<[E]%
\\
\>[3]{}\hsindent{2}{}\<[5]%
\>[5]{}\mathbin{::}\forall \Varid{insS}\hsforall \codeskip \Varid{insT}\hsdot{\cdot }{.}(\Conid{InitialPipes}\codeskip \Varid{insS}\codeskip \Varid{insT})\Rightarrow \Conid{IO}\codeskip (\Conid{BasicNetwork}\codeskip \Varid{insS}\codeskip \Varid{insT}\codeskip \Varid{insS}\codeskip \Varid{insT}){}\<[E]%
\\
\>[3]{}\Varid{initialNetwork}\mathrel{=}\mathbf{do}{}\<[E]%
\\
\>[3]{}\hsindent{2}{}\<[5]%
\>[5]{}\Varid{ps}\leftarrow \Varid{initialPipes}\mathbin{::}\Conid{IO}\codeskip (\Conid{PipeList}\codeskip \Varid{insS}\codeskip \Varid{insT}){}\<[E]%
\\
\>[3]{}\hsindent{2}{}\<[5]%
\>[5]{}\keyw{return}\codeskip (\Conid{BasicNetwork}\codeskip \Varid{empty}\codeskip \Varid{empty}\codeskip \Varid{ps}\codeskip \Varid{ps}){}\<[E]%
\ColumnHook
\end{hscode}\resethooks
The \ensuremath{\Conid{InitialPipes}} type class constructs an \ensuremath{\Varid{initialPipes}} based on the type required in the initial network.

\subsubsection{The Translation}
Now that the algebra type class, and the initial input to the accumulating fold is defined, each instance of the type class can be defined.

\paragraph{Basic Constructors}
There are several constructors that just manipulate the output \ensuremath{\Conid{PipeList}}, these constructors are \ensuremath{\Conid{Id}}, \ensuremath{\Conid{Replicate}}, \ensuremath{\Conid{Swap}}, \ensuremath{\Conid{DropL}}, and \ensuremath{\Conid{DropR}}.
The \ensuremath{\Conid{Swap}} constructor takes two inputs and then swaps them over:
\begin{hscode}\SaveRestoreHook
\column{B}{@{}>{\hspre}l<{\hspost}@{}}%
\column{3}{@{}>{\hspre}l<{\hspost}@{}}%
\column{5}{@{}>{\hspre}l<{\hspost}@{}}%
\column{9}{@{}>{\hspre}l<{\hspost}@{}}%
\column{20}{@{}>{\hspre}l<{\hspost}@{}}%
\column{E}{@{}>{\hspre}l<{\hspost}@{}}%
\>[3]{}\mathbf{instance}\codeskip \Conid{BuildNetworkAlg}\codeskip \Conid{BasicNetwork}\codeskip \Conid{Swap}\codeskip \mathbf{where}{}\<[E]%
\\
\>[3]{}\hsindent{2}{}\<[5]%
\>[5]{}\Varid{buildNetworkAlg}\codeskip \Conid{Swap}\mathrel{=}\keyw{return}\mathbin{\$}\Conid{AccuN}\codeskip (\lambda \Varid{n}\to \mathbf{do}{}\<[E]%
\\
\>[5]{}\hsindent{4}{}\<[9]%
\>[9]{}\mathbf{let}\codeskip \Conid{PipeCons}\codeskip \Varid{c1}\codeskip (\Conid{PipeCons}\codeskip \Varid{c2}\codeskip \Conid{PipeNil})\mathrel{=}\Varid{outs}\codeskip \Varid{n}{}\<[E]%
\\
\>[5]{}\hsindent{4}{}\<[9]%
\>[9]{}\keyw{return}\mathbin{\$}\Conid{BasicNetwork}\codeskip {}\<[E]%
\\
\>[9]{}\hsindent{11}{}\<[20]%
\>[20]{}(\Varid{threads}\codeskip \Varid{n})\codeskip (\Varid{jobs}\codeskip \Varid{n})\codeskip (\Varid{ins}\codeskip \Varid{n})\codeskip {}\<[E]%
\\
\>[9]{}\hsindent{11}{}\<[20]%
\>[20]{}(\Conid{PipeCons}\codeskip \Varid{c2}\codeskip (\Conid{PipeCons}\codeskip \Varid{c1}\codeskip \Conid{PipeNil}))){}\<[E]%
\ColumnHook
\end{hscode}\resethooks

The instance for \ensuremath{\Conid{Swap}}, defines a function wrapped by \ensuremath{\Conid{AccuN}}, that takes the current accumulated network, up to this point.
It then transforms the outputs by swapping \ensuremath{\Varid{c1}} and \ensuremath{\Varid{c2}}, and building a new \ensuremath{\Conid{BasicNetwork}}.
All other leaf constructors will follow this pattern.

\paragraph{Task}
In a \ensuremath{\Conid{BasicNetwork}}, a task will run as a separate thread, to do this \ensuremath{\Varid{forkIO}\mathbin{::}\Conid{IO}\codeskip ()\to \Conid{IO}\codeskip \Conid{ThreadId}} will be used.
Using this function requires some \ensuremath{\Conid{IO}\codeskip ()} computation to run, this will be defined by \ensuremath{\Varid{taskExecutor}}, which will read a value from each of input channels, execute the task with those inputs, and then write the output to the output channels.
This computation is then repeated forever. Making use of the \ensuremath{\Varid{taskExecutor}}, the algebra instance for \ensuremath{\Conid{Task}} is as:
\begin{hscode}\SaveRestoreHook
\column{B}{@{}>{\hspre}l<{\hspost}@{}}%
\column{3}{@{}>{\hspre}l<{\hspost}@{}}%
\column{5}{@{}>{\hspre}l<{\hspost}@{}}%
\column{9}{@{}>{\hspre}l<{\hspost}@{}}%
\column{19}{@{}>{\hspre}l<{\hspost}@{}}%
\column{20}{@{}>{\hspre}l<{\hspost}@{}}%
\column{48}{@{}>{\hspre}l<{\hspost}@{}}%
\column{E}{@{}>{\hspre}l<{\hspost}@{}}%
\>[3]{}\mathbf{instance}\codeskip \Conid{BuildNetworkAlg}\codeskip \Conid{BasicNetwork}\codeskip \Conid{Task}\codeskip \mathbf{where}{}\<[E]%
\\
\>[3]{}\hsindent{2}{}\<[5]%
\>[5]{}\Varid{buildNetworkAlg}\codeskip (\Conid{Task}\codeskip \Varid{t})\mathrel{=}\keyw{return}\mathbin{\$}\Conid{AccuN}\codeskip (\lambda \Varid{n}\to \mathbf{do}{}\<[E]%
\\
\>[5]{}\hsindent{4}{}\<[9]%
\>[9]{}\Varid{out}{}\<[19]%
\>[19]{}\leftarrow \Conid{PipeCons}<\!\!\$\!\!>\Varid{newChan}<\!\!*\!\!>{}\<[48]%
\>[48]{}\keyw{return}\codeskip \Conid{PipeNil}{}\<[E]%
\\
\>[5]{}\hsindent{4}{}\<[9]%
\>[9]{}\Varid{taskUUID}{}\<[19]%
\>[19]{}\leftarrow \Varid{genUnusedTaskUUID}\codeskip (\Varid{threads}\codeskip \Varid{n}){}\<[E]%
\\
\>[5]{}\hsindent{4}{}\<[9]%
\>[9]{}\Varid{threadId}{}\<[19]%
\>[19]{}\leftarrow \Varid{forkIO}\codeskip (\Varid{taskExecuter}\codeskip (\Conid{Task}\codeskip \Varid{t})\codeskip \Varid{taskUUID}\codeskip (\Varid{outputs}\codeskip \Varid{n})\codeskip \Varid{output}){}\<[E]%
\\
\>[5]{}\hsindent{4}{}\<[9]%
\>[9]{}\keyw{return}\mathbin{\$}\Conid{BasicNetwork}\codeskip {}\<[E]%
\\
\>[9]{}\hsindent{11}{}\<[20]%
\>[20]{}(\Varid{\Conid{M}.insert}\codeskip \Varid{taskUUID}\codeskip \Varid{threadId}\codeskip (\Varid{threads}\codeskip \Varid{n}))\codeskip (\Varid{jobs}\codeskip \Varid{n})\codeskip (\Varid{inputs}\codeskip \Varid{n})\codeskip \Varid{output}{}\<[E]%
\ColumnHook
\end{hscode}\resethooks
This instance first creates a new output channel, this will be given to the task to send its outputs on.
It then forks a new thread with the computation generated by \ensuremath{\Varid{taskExecutor}}.
The executor is given the output values of the accumulated network and the output channel, just created.
The resulting network has the same inputs, but now adds a new thread id to the list and the outputs set to be the output channels from the task.

\paragraph{Then}
The \ensuremath{\Conid{Then}} constructor is responsible for connecting circuits in sequence.
When converting this to a network, this will involve making use of the accumulated network value to generate the next layer.
The instance is defined as:
\begin{hscode}\SaveRestoreHook
\column{B}{@{}>{\hspre}l<{\hspost}@{}}%
\column{3}{@{}>{\hspre}l<{\hspost}@{}}%
\column{5}{@{}>{\hspre}l<{\hspost}@{}}%
\column{7}{@{}>{\hspre}l<{\hspost}@{}}%
\column{E}{@{}>{\hspre}l<{\hspost}@{}}%
\>[3]{}\mathbf{instance}\codeskip \Conid{BuildNetworkAlg}\codeskip \Conid{BasicNetwork}\codeskip \Conid{Then}\codeskip \mathbf{where}{}\<[E]%
\\
\>[3]{}\hsindent{2}{}\<[5]%
\>[5]{}\Varid{buildNetworkAlg}\codeskip (\Conid{Then}\codeskip (\Conid{AccuN}\codeskip \Varid{fx})\codeskip (\Conid{AccuN}\codeskip \Varid{fy})){}\<[E]%
\\
\>[5]{}\hsindent{2}{}\<[7]%
\>[7]{}\mathrel{=}\keyw{return}\mathbin{\$}\Conid{AccuN}\codeskip (\Varid{fx}\mathbin{>=>}\Varid{fy}){}\<[E]%
\ColumnHook
\end{hscode}\resethooks
This instance has an interesting definition: firstly it takes the accumulated network \ensuremath{\Varid{n}} as input.
It then uses the function \ensuremath{\Varid{fx}}, with the input \ensuremath{\Varid{n}} to generate a network for the top half of the \ensuremath{\Conid{Then}} constructor.
Finally, it takes the returned network, from the top half of the constructor, and generates a network using the function \ensuremath{\Varid{fy}} representing the bottom half of the constructor.

\paragraph{Beside} The \ensuremath{\Conid{Beside}} (\ensuremath{\mathbin{<>}}) constructor places two circuits side by side.
This is the most tedious algebra to define as the accumulated network needs to be split in half to pass to the two recursive sides of \ensuremath{\Conid{Beside}}. Details of its translation can be found in Appendix E.

\ensuremath{\Conid{CircuitFlow}} also uses the \ensuremath{\Conid{ExceptT}} monad transformer to fail gracefully.
\section{Benchmarks}\label{sec:benchmarks}

We use the audio streaming example from Section~\ref{sec:lang} to perform the benchmarking.
It is also the main application domain of Luigi which we will compare with.
Haskell benchmarks were taken using \ensuremath{\Varid{criterion}}~\cite{criterion}; Python 3.8.5 benchmarks with \ensuremath{\Varid{pytest}\mathbin{-}\Varid{benchmark}}~\cite{pytime}.
Each benchmark is tested on thirteen different numbers of inputs: 1, 10, 100, 200,
then at intervals of 200 until 2000, with measurements repeated and summarised as a mean average.

Three months of one of the author's own audio history is used, to ensure that the data closely aligns with the real world.
This allows for the evaluation of how each implementation scales with more inputs.
All benchmarks take place on an Intel(R) Core(TM) i5-4690 CPU at 3.50GHz (4 cores and no hyper-threading), with 8GB of RAM booting Ubuntu 20.04.

\subsubsection{Multi-Core Haskell}
By default the Haskell runtime does not enable multi-core processing.
Considering the aim of this project partly involves making \ensuremath{\Conid{CircuitFlow}} run in parallel, multi-core processing is crucial.
To enable this the \texttt{-threaded} flag is set when building the binary.
Then, using the runtime options, the number of threads can be set by adding \texttt{+RTS -N} flags when running the binary.
The \texttt{-N} allows the runtime to select the optimal number of threads for the program.

\subsubsection{Parallel vs Serial}
The first test will ensure that \ensuremath{\Conid{CircuitFlow}}'s parallelisation has a positive effect on run-times.
To ensure that the test is fair, the serial implementation will make use of the same tasks in the pre-processing pipeline.
The inputs and outputs will just be manually fed into each task, in a sequential way.
The results show that \ensuremath{\Conid{CircuitFlow}} does indeed provide a performance gain, with a mean speedup of 1.53x.

Profiling the circuit shows that a significant proportion of time is spend reading CSV files. Optimising speed of CSV parsing and how often a CSV is read via caching would improve runtime. Another area for improvement is that there is an expectation on the user to know where is best to split up the workflow into tasks.
It would be beneficial if a circuit could automatically fuse tasks together,
then it would have a positive effect on the runtime.

\begin{figure}
     \hspace{0.2\textwidth}
     \begin{subfigure}[b]{0.2\textwidth}
        \centering
        \abovecaptionskip
         \belowcaptionskip
        \begin{tikzpicture}[scale=0.4]

          \definecolor{color0}{rgb}{0,0,0.0502}
          \definecolor{color1}{rgb}{1,0.498039215686275,0.0549019607843137}

          \begin{axis}[
          legend cell align={left},
          legend style={
            fill opacity=0.8,
            draw opacity=1,
            text opacity=1,
            at={(0.03,0.97)},
            anchor=north west,
            draw=white!80!black
          },
          tick align=outside,
          tick pos=left,
          x grid style={white!69.0196078431373!black},
          xlabel={Number of inputs},
          xmin=-98.95, xmax=2099.95,
          xtick style={color=black},
          y grid style={white!69.0196078431373!black},
          ylabel={Runtime (s)},
          ymin=-13.689405, ymax=290.423305,
          ytick style={color=black}
          ]
          \addplot [semithick, color0, mark=x, mark size=3, mark options={solid}]
          table {%
          1 0.1349
          10 1.36
          100 13.61
          200 27.18
          400 54.74
          600 82.43
          800 109.8
          1000 137.4
          1200 165
          1400 192.1
          1600 220.3
          1800 248.9
          2000 276.6
          };
          \addlegendentry{Single-core CircuitFlow}
          \addplot [semithick, color1, mark=x, mark size=3, mark options={solid}]
          table {%
          1 0.1339
          10 1.336
          100 13.31
          200 26.62
          400 53.14
          600 79.92
          800 106.6
          1000 133
          1200 159.9
          1400 186.5
          1600 212.8
          1800 239.6
          2000 266.4
          };
          \addlegendentry{Serial}
          \end{axis}
        \end{tikzpicture}
        \caption{ Linear}
        \label{fig:linear-linear-c}
     \end{subfigure}
     \hspace{0.1\textwidth}
     \begin{subfigure}[b]{0.2\textwidth}
        \centering
        \abovecaptionskip
         \belowcaptionskip
        \begin{tikzpicture}[scale=0.4]

          \definecolor{color0}{rgb}{0,0,0.0502}
          \definecolor{color1}{rgb}{0.2980,0.73333,0.090196} 

          \begin{axis}[
          legend cell align={left},
          legend style={
            fill opacity=0.8,
            draw opacity=1,
            text opacity=1,
            at={(0.03,0.97)},
            anchor=north west,
            draw=white!80!black
          },
          tick align=outside,
          tick pos=left,
          x grid style={white!69.0196078431373!black},
          xlabel={Number of inputs},
          xmin=-98.95, xmax=2099.95,
          xtick style={color=black},
          y grid style={white!69.0196078431373!black},
          ylabel={Runtime (s)},
          ymin=-35.068680885, ymax=738.317578585,
          ytick style={color=black}
          ]
          \addplot [semithick, color0, mark=x, mark size=3, mark options={solid}]
          table {%
          1 0.08524
          10 0.8553
          100 8.711
          200 17.38
          400 34.74
          600 52.15
          800 70.2
          1000 86.66
          1200 104.9
          1400 122.3
          1600 140.4
          1800 157.9
          2000 175.5
          };
          \addlegendentry{CircuitFlow}
          \addplot [semithick, color1, mark=x, mark size=3, mark options={solid}]
          table {%
          1 0.0990363
          10 0.5708444
          100 7.1214412
          200 13.5843029
          400 36.9573301
          600 77.2875785
          800 127.2281325
          1000 190.2836947
          1200 267.5381546
          1400 357.0366452
          1600 462.572336
          1800 572.1155158
          2000 703.1636577
          };
          \addlegendentry{Luigi}
          \end{axis}

        \end{tikzpicture}

        \caption{Vs Luigi}
        \label{fig:luigi-parallel-c}
     \end{subfigure}
        \caption{\ensuremath{\Conid{CircuitFlow}} benchmarks}
        \abovecaptionskip
         \belowcaptionskip
        \label{fig:three graphs}
\end{figure}

\subsubsection{1 Core Circuit vs Serial}
Another interesting scenario to test is checking if the network structure adds additional overhead, in a situation where there is only 1 core.
To test this, the multi-core support of the Haskell runtime will not be enabled: this will then simulate multiple cores with context switching.
Figure~\ref{fig:linear-linear-c}, shows the results of this benchmark. It shows that both the linear and single core implementation scale together in a linear fashion.
Most importantly, \ensuremath{\Conid{CircuitFlow}} only adds a minor overhead over a linear implementation.
This will be particularly helpful for a user which needs to run code on multiple types of devices.
There is no need for them to create a different implementation for devices where parallelisation may not be possible.

\subsubsection{\ensuremath{\Conid{CircuitFlow}} vs Luigi}
The final benchmark on \ensuremath{\Conid{CircuitFlow}} is comparing it to widely used library: Luigi by Spotify~\cite{spotify_luigi}.
Since Luigi uses a \ac{DPN}, it can use any number of threads: in this test it is set to 4 --- the same as \ensuremath{\Conid{CircuitFlow}}.
Figure~\ref{fig:luigi-parallel-c}, shows the results of the benchmark.

This shows that \ensuremath{\Conid{CircuitFlow}} performs better than Luigi on larger numbers of inputs.
\ensuremath{\Conid{CircuitFlow}} scales linearly with the number of inputs, whereas Luigi's runtime appears to grow at a quicker rate than linear.

\subsubsection{Why is \ensuremath{\Conid{CircuitFlow}} so good?}
Luigi and \ensuremath{\Conid{CircuitFlow}} have their differences, which will likely explain why there is a difference in run times, especially with larger numbers of inputs.

\paragraph{More Lightweight}
Luigi is a far more complex library with advanced features, not included in \ensuremath{\Conid{CircuitFlow}}, that may slow Luigi down --- one such feature is back filling.
This allows Luigi to avoid running tasks that have already been run.
This feature means that before executing a task the Luigi scheduler has to check if a task has already been executed.
This adds additional overhead to the scheduler that \ensuremath{\Conid{CircuitFlow}} does not have.
Although this feature does have its benefits, after the first run of Luigi all run times after are very quick as no tasks will need to be executed.
If \ensuremath{\Conid{CircuitFlow}} were to implement this feature any overhead it adds will be partially mitigated by the checks being distributed across multiple threads, instead of in one central scheduler.

\paragraph{Computation Models}
The two libraries use variants of the same computation model: \ensuremath{\Conid{CircuitFlow}} uses a KPN and Luigi uses a \ac{DPN}~\cite{381846}.
This difference is the main reason why \ensuremath{\Conid{CircuitFlow}} scales linearly when it needs to process more input values.
\ensuremath{\Conid{CircuitFlow}} makes use of buffered channels to keep a queue of all inputs that need to be processed.
However, Luigi does not rely on this design, instead it has a pool of workers with a scheduler controlling what is executed on each worker.
It is this scheduler that causes Luigi to scale non-linearly.
As the number of inputs grow, the scheduler will have to schedule more and more tasks: this process is not $\mathcal{O}(n)$.

\paragraph{Multi-processing in Python}
\ensuremath{\Conid{CircuitFlow}} makes use of a static number of threads defined by the number of tasks in a circuit.
Luigi on the other hand can support any number of workers, however, Luigi suffers from a downfall of Python: threads cannot run in parallel due to the Global Interpreter Lock.
To avoid this Luigi uses processes not threads, which adds extra overhead.
Luigi also creates a new process for each invocation of a task, which \ensuremath{\Conid{CircuitFlow}} does not do.
This means that Luigi will start 8000 processes vs \ensuremath{\Conid{CircuitFlow}}'s 4 threads for the 2000 inputs benchmark.
\ensuremath{\Conid{CircuitFlow}}'s static number of threads could also be considered a downside due to the lack of flexibility depending on run-time values.
To combat this more combinators can be introduced that allow for branching or other similar operations, in fact, \ensuremath{\Varid{mapC}} is a combinator of this type.

\section{Discussion and Related Work}\label{sec:relatedwork}

In this section, we cover the embedding techniques that we build upon and how our process can be replicated.
We also discuss other popular workflow libraries including imperative and functional ones, comparing them to \ensuremath{\Conid{CircuitFlow}}.


\paragraph{Summary of Embedding Techniques and their General Use.}
\ensuremath{\Conid{CircuitFlow}}, which can be more generally be seen as an \ac{eDSL} whose semantics needs to use effects and has rich types to verify program correctness, has been created in a modular manner that doesn't compromise on performance.
The pivotal parts of \ensuremath{\Conid{CircuitFlow}}'s creation can be replicated to produce such other \acp{eDSL}.
The process is one of three parts.
The first is the curation of the type information.
In the case of \ensuremath{\Conid{CircuitFlow}}, this was dependency information, and was achieved using Haskell's approximation of dependant types (DataKinds~\cite{10.1145/2103786.2103795} and Singletons~\cite{10.1145/2364506.2364522} for value promotion/demotion to/from the type level; Type Families~\cite{10.1145/1411204.1411215} for type information manipulation; and Heterogeneous lists~\cite{10.1145/1017472.1017488} for, well, storing more than one type in a list).
The second act follows the same beats as the classic embedding story~\cite{foldingDSLs}: each construct is created as a separate fixed functor, where all constructs can be composed together with the beloved \textit{Data types \`{a} la carte}, and semantics provided through a ``classy'' algebra.
The story just needs to be jacked up to accommodate the type information and effects, with the trick being the switch to indexed functors~\cite{mcbride2011functional,10.1145/1328438.1328475} and a monadic catamorphism~\cite{monadic_cata}.
Finally, the choice of underlying semantics is key for speed as that is ultimately what will be running. Our choice of \acp{KPN} assisted us greatly with \ensuremath{\Conid{CircuitFlow}}'s competitive run-time.

\paragraph{Applicative Functors}
An example of capturing parallelism in Haskell is to use applicative functors~\cite{10.1017/S0956796807006326} --- a technique employed by the Haxl library~\cite{10.1145/2628136.2628144}.
This approach can leverage the applicative combinators to group together computation that can be performed simultaneously.
There is even the \ensuremath{\Conid{ApplicativeDo}} language extension~\cite{10.1145/3241625.2976007}, which desugars do notation down to applicative combinators.
However, this approach suffers from some forced sequentiality at points.
Take the previously mentioned example Figure \ref{fig:example-song-pre-proc}, both top ten tasks would be grouped together.
Leading to neither task being able to begin until both aggregations have been completed.

\paragraph{Arrows}
Another method is arrows~\cite{HUGHES200067}, used by Funflow based on \textit{Composing Effects into Tasks and Workflows}~\cite{10.1145/3406088.3409023}.
Arrows similarly are often used through with the notation obtained from the language extension~\cite{PatersonRA:notation}, which introduces a do style notation.
They also fall victim to the same problems as applicative functors.
Due to the constructor \ensuremath{\Varid{arr}} consuming a function, it is not possible to inspect inside and fully exploit all cases of parallelisation.

The Funflow library that makes use of arrows, does so by noticing that tasks in workflows are similar to effects in the functional community.
It draws from existing work on combining and analysing effects, with categories and arrows, and applies this to constructing workflows.

\paragraph{\acp{SMC}}
Linear Haskell is put to excellent use in \textit{Evaluating Linear Functions to Symmetric Monoidal Categories}~\cite{10.1145/3471874.3472980} to address the problem of over sequentialisation found in applicative functors and arrows.
It introduces a new \ac{SMC} type class that allows for all parallelism to be exposed and exploited in a workflow.
The type class adds new combinators for linear Haskell functions, that can be composed in a style that aligns with do notation.
It uses atomic types to detail the synchronisation points, and where synchronisation can be discovered by a scheduler.
However, it comes with the caveat that it can only compose linear functions.

\paragraph{Pipes~\cite{pipes}} focuses on supporting steaming data, which is beneficial as there is no need to wait for jobs to finish before moving on. This is something that \ensuremath{\Conid{CircuitFlow}} is also designed to support without any modifications: a network can be started and inputs can be streamed in when they are available.

\paragraph{Luigi~\cite{spotify_luigi}}
Industry-favourite Luigi, used to orchestrate tasks in a data workflow, is a library that, as we have seen, falls into the trap of un-typed task dependencies.
It makes use of a central scheduler and workers, allowing work to be distributed across multiple machines.
It also comes with built-in support for many different output formats, such as files in a Hadoop file system.

\paragraph{SciPipe~\cite{10.1093/gigascience/giz044}}
An approach for orchestrating external jobs is taken by SciPipe, a workflow library for agile development of complex and dynamic bioinformatics pipelines. Unlike \ensuremath{\Conid{CircuitFlow}} and many other libraries, instead of defining tasks as functions within the embedding language, SciPipe uses Bash commands to easily interact with pre-existing binaries. This allows task to be written in the language most suited for its requirements, however, comes with the downside of the additional infrastructure required to create all these binaries for each task. Due to the separation of tasks into bash scripts, type checking interactions between tasks is significantly harder.

\paragraph{Other Typed Dataflow Libraries}
DryadLINQ~\cite{10.5555/1855741.1855742} allows for developers to create parallel programs in SQL-like LINQ expressions. Similarly to \ensuremath{\Conid{CircuitFlow}}, these can be inspected to find any data-parallel sections and then automatically translated into a distributed execution plan that can run on Dryad --- although \ensuremath{\Conid{CircuitFlow}} currently lacks a distributed network implementation.
FlumeJava~\cite{35650}, uses lazy evaluation of operations on parallel data structures, to build a dataflow graph of the steps required. When the value is required the graph is optimised to evaluate the operations in an optimal way.
Unlike \ensuremath{\Conid{CircuitFlow}}, Naiad~\cite{murray2013naiad} can execute cyclic dataflow programs. It does so on a distributed system, to help with streaming data analysis or iterative machine learning training.

\paragraph{Staged Selective Parser Combinators~\cite{parsley}}
Indexed functors~\cite{mcbride2011functional}, are a new technique for building typed \ac{eDSL}.
This paper makes use of this new tool to have a type index representing the type of a parser.
This allows it to make optimisations and translations while ensuring that the value parsed never changes.
\section{Conclusion}
This paper introduced a new \ac{eDSL} to declaratively construct data workflows, which are type-safe and competitive in run-time performance.
The design of \ensuremath{\Conid{CircuitFlow}} draws its origins from a strong mathematical background, with each constructor directly representing an axiom in a \ac{SMP}.
This demonstrates the language's completeness at being able to represent any \ac{DAG}, that a data workflow may need.
The battle for type-safety without compromising run-time or modular design was a tough one, but one that can be replicated to great avail when creating languages with a similar requirements.

\subsubsection*{Acknowledgements.}
The authors would like to thank Jamie Willis for his insights while creating \ensuremath{\Conid{CircuitFlow}} and the anonymous reviewers for their constructive and helpful comments.

The work is partly supported by EPSRC Grant \emph{EXHIBIT: Expressive
High-Level Languages for Bidirectional Transformations} (EP/T008911/1)
and Royal Society Grant \emph{Bidirectional Compiler for Software Evolution} (IESR3170104).


\noindent
\bibliographystyle{../splncs04}
\bibliography{../references}

\chapter*{Appendices}








\appendix

\section{Type Heft}\label{sec:type}

\ensuremath{\Conid{CircuitFlow}} is crafted using a good proportion of the type heft that Haskell can muster. Herein lies the reason that creating a standalone language was simply out of the question.

This section provides a quick rundown of the Haskell type magic used to create \ensuremath{\Conid{CircuitFlow}}.

\subsection{Dependently Typed Programming}\label{sec:bg-dependently-typed}

Although Haskell does not officially support dependently typed programming, there are techniques available that together can be used to replicate some of the experience.

\begin{itemize}
  \item DataKinds~\cite{10.1145/2103786.2103795} allows for values to be promoted to types. When constructors are promoted to type constructors, they are prefixed with a \ensuremath{'} . This allows for more interesting and restrictive types such as vectors that store their length at the type level.

  \item Singletons~\cite{10.1145/2364506.2364522} allow types to be demoted to values. They are called singletons because they only have one inhabitant - one possible value for each type. This way the constructed type can reflect its type back to the value level. An example usage is recovering the length of a vector as a singleton \ensuremath{\Conid{SNat}} value.

  \item Type Families~\cite{10.1145/1411204.1411215} can be used to define functions that manipulate types. Type level functions are defined just like normal Haskell functions: by pattern matching on the constructors, they just have slightly different syntax.

  \item Heterogeneous lists~\cite{10.1145/1017472.1017488} leverage the promotion of values to the type level to allow for lists of multiple types.
  Rather than be parameterised by a single type, they instead make use of a type list, which is the list type promoted through DataKinds to be a kind, with its elements being types. Each element in the type list aligns with the value at that position in the list, giving its type. A heterogeneous list is defined as:
\begin{hscode}\SaveRestoreHook
\column{B}{@{}>{\hspre}l<{\hspost}@{}}%
\column{3}{@{}>{\hspre}l<{\hspost}@{}}%
\column{5}{@{}>{\hspre}l<{\hspost}@{}}%
\column{12}{@{}>{\hspre}l<{\hspost}@{}}%
\column{E}{@{}>{\hspre}l<{\hspost}@{}}%
\>[3]{}\mathbf{data}\codeskip \Conid{HList}\codeskip (\Varid{xs}\mathbin{::}[\mskip1.5mu \Conid{Type}\mskip1.5mu])\codeskip \mathbf{where}{}\<[E]%
\\
\>[3]{}\hsindent{2}{}\<[5]%
\>[5]{}\Conid{HNil}{}\<[12]%
\>[12]{}\mathbin{::}\Conid{HList}\codeskip '[\mskip1.5mu \mskip1.5mu]{}\<[E]%
\\
\>[3]{}\hsindent{2}{}\<[5]%
\>[5]{}\Conid{HCons}{}\<[12]%
\>[12]{}\mathbin{::}\Varid{x}\to \Conid{HList}\codeskip \Varid{xs}\to \Conid{HList}\codeskip (\Varid{x}\codeskip \,'\!\!\mathbin{:}\codeskip \Varid{xs}){}\<[E]%
\ColumnHook
\end{hscode}\resethooks
\end{itemize}

\subsection{Phantom Types}

Phantom type parameters~\cite{phantom_types} are when a type variable only appears on the left hand side of the equals.
The most basic example is \ensuremath{\Conid{Const}}, it has two type arguments, but only \ensuremath{\Varid{a}} is used on the right hand side. Phantom type parameters can be used to store information in the types, which can act as further static constraints on the types.

\section{Build System (lhs2TeX)}\label{sec:build_example}

Another use case for a task based dependency system is a build system.
For example, a Makefile is a way of specifying the target files from some source files, with a command that can be used to generate the target file.
\ensuremath{\Conid{CircuitFlow}}: could also be used to model such a system.

Consider this paper, which is made using \LaTeX.
This project is made up of multiple subfiles, each written in a literate Haskell format.
Each of these files needs to be pre-processed by the \texttt{lhs2TeX} command to produce the \texttt{.tex} source file.
Once each of these files has been generated, then the \LaTeX project can be built into a PDF file.

\subsection{Building the Circuit}
The \ensuremath{\Conid{Circuit}} defined here makes use of the \ensuremath{\Varid{mapC}} operator.
To do so a \ensuremath{\Conid{Circuit}} is defined that is able to build a single \texttt{.tex} file from a \texttt{.lhs}.
This has to make use of the standard \ensuremath{\Varid{task}} constructor:

\noindent\begin{minipage}{\linewidth}
\begin{hscode}\SaveRestoreHook
\column{B}{@{}>{\hspre}l<{\hspost}@{}}%
\column{3}{@{}>{\hspre}l<{\hspost}@{}}%
\column{5}{@{}>{\hspre}l<{\hspost}@{}}%
\column{9}{@{}>{\hspre}l<{\hspost}@{}}%
\column{E}{@{}>{\hspre}l<{\hspost}@{}}%
\>[B]{}\Varid{lhs2TexTask}\mathbin{::}\Conid{Circuit}\codeskip '[\mskip1.5mu \Conid{Var}\mskip1.5mu]\codeskip '[\mskip1.5mu \Conid{String}\mskip1.5mu]\codeskip '[\mskip1.5mu \Conid{Var}\mskip1.5mu]\codeskip '[\mskip1.5mu \Conid{String}\mskip1.5mu]\codeskip \Conid{N1}{}\<[E]%
\\
\>[B]{}\Varid{lhs2TexTask}\mathrel{=}\Varid{task}\codeskip \Varid{f}{}\<[E]%
\\
\>[B]{}\mathbf{where}{}\<[E]%
\\
\>[B]{}\hsindent{3}{}\<[3]%
\>[3]{}\Varid{f}\mathbin{::}\Conid{IHList}\codeskip '[\mskip1.5mu \Conid{Var}\mskip1.5mu]\codeskip '[\mskip1.5mu \Conid{String}\mskip1.5mu]\to \Conid{Var}\codeskip \Conid{String}\to \Conid{ExceptT}\codeskip \Conid{SomeException}\codeskip \Conid{IO}\codeskip (){}\<[E]%
\\
\>[B]{}\hsindent{3}{}\<[3]%
\>[3]{}\Varid{f}\codeskip \Varid{input}\codeskip \Varid{output}\mathrel{=}\Varid{lift}\codeskip (\mathbf{do}{}\<[E]%
\\
\>[3]{}\hsindent{2}{}\<[5]%
\>[5]{}\Conid{HCons}\codeskip \Varid{fInName}\codeskip \Conid{HNil}\leftarrow \Varid{fetch'}\codeskip \Varid{input}{}\<[E]%
\\
\>[3]{}\hsindent{2}{}\<[5]%
\>[5]{}\mathbf{let}\codeskip \Varid{fOutName}\mathrel{=}\Varid{fInName}-\!\!<\!\!.\!\!>\text{\ttfamily \char34 tex\char34}{}\<[E]%
\\
\>[3]{}\hsindent{2}{}\<[5]%
\>[5]{}\Varid{callCommand}\codeskip {}\<[E]%
\\
\>[5]{}\hsindent{4}{}\<[9]%
\>[9]{}(\text{\ttfamily \char34 lhs2tex~-o~\char34}\plus \Varid{fOutName}\plus \text{\ttfamily \char34 ~\char34}\plus \Varid{fInName}\plus \text{\ttfamily \char34 ~>~lhs2tex.log\char34}){}\<[E]%
\\
\>[3]{}\hsindent{2}{}\<[5]%
\>[5]{}\Varid{save}\codeskip \Varid{output}\codeskip \Varid{fOutName}){}\<[E]%
\ColumnHook
\end{hscode}\resethooks
\end{minipage}

This \ensuremath{\Conid{Circuit}} makes use of the \ensuremath{\Varid{callCommand}} function from the \ensuremath{\Conid{\Conid{System}.Process}} library.
This allows the task to execute external commands, that may not necessarily be defined in Haskell.
A similar \ensuremath{\Conid{Circuit}} can be defined that will compile the \texttt{.tex} files and produce a PDF.

\noindent\begin{minipage}{\linewidth}
\begin{hscode}\SaveRestoreHook
\column{B}{@{}>{\hspre}l<{\hspost}@{}}%
\column{3}{@{}>{\hspre}l<{\hspost}@{}}%
\column{5}{@{}>{\hspre}l<{\hspost}@{}}%
\column{9}{@{}>{\hspre}l<{\hspost}@{}}%
\column{E}{@{}>{\hspre}l<{\hspost}@{}}%
\>[B]{}\Varid{buildTexTask}\mathbin{::}\Conid{String}\to \Conid{String}\to \Conid{Circuit}\codeskip '[\mskip1.5mu \Conid{Var}\mskip1.5mu]\codeskip '[\mskip1.5mu [\mskip1.5mu \Conid{String}\mskip1.5mu]\mskip1.5mu]\codeskip '[\mskip1.5mu \Conid{Var}\mskip1.5mu]\codeskip '[\mskip1.5mu \Conid{String}\mskip1.5mu]\codeskip \Conid{N1}{}\<[E]%
\\
\>[B]{}\Varid{buildTexTask}\codeskip \Varid{outputFileName}\codeskip \Varid{mainFileName}\mathrel{=}\Varid{task}\codeskip \Varid{f}{}\<[E]%
\\
\>[B]{}\mathbf{where}{}\<[E]%
\\
\>[B]{}\hsindent{3}{}\<[3]%
\>[3]{}\Varid{f}\mathbin{::}\Conid{IHList}\codeskip '[\mskip1.5mu \Conid{Var}\mskip1.5mu]\codeskip '[\mskip1.5mu [\mskip1.5mu \Conid{String}\mskip1.5mu]\mskip1.5mu]\to \Conid{Var}\codeskip \Conid{String}\to \Conid{ExceptT}\codeskip \Conid{SomeException}\codeskip \Conid{IO}\codeskip (){}\<[E]%
\\
\>[B]{}\hsindent{3}{}\<[3]%
\>[3]{}\Varid{f}\codeskip \anonymous \codeskip \Varid{output}\mathrel{=}\Varid{lift}\codeskip (\mathbf{do}{}\<[E]%
\\
\>[3]{}\hsindent{2}{}\<[5]%
\>[5]{}\Varid{callCommand}\codeskip {}\<[E]%
\\
\>[5]{}\hsindent{4}{}\<[9]%
\>[9]{}(\text{\ttfamily \char34 texfot~--no-stderr~latexmk~-interaction=nonstopmode~\char34}{}\<[E]%
\\
\>[5]{}\hsindent{4}{}\<[9]%
\>[9]{}\plus \text{\ttfamily \char34 -pdf~-no-shell-escape~-bibtex~-jobname=\char34}{}\<[E]%
\\
\>[5]{}\hsindent{4}{}\<[9]%
\>[9]{}\plus \Varid{outputFileName}{}\<[E]%
\\
\>[5]{}\hsindent{4}{}\<[9]%
\>[9]{}\plus \text{\ttfamily \char34 ~\char34}{}\<[E]%
\\
\>[5]{}\hsindent{4}{}\<[9]%
\>[9]{}\plus \Varid{mainFileName}{}\<[E]%
\\
\>[5]{}\hsindent{4}{}\<[9]%
\>[9]{}){}\<[E]%
\\
\>[3]{}\hsindent{2}{}\<[5]%
\>[5]{}\Varid{save}\codeskip \Varid{output}\codeskip (\Varid{outputFileName}<\!\!.\!\!>\text{\ttfamily \char34 pdf\char34})){}\<[E]%
\ColumnHook
\end{hscode}\resethooks
\end{minipage}

The \ensuremath{\Varid{buildTexTask}} also demonstrates how it is possible to pass a global parameter into a task.
This can allow tasks to be made in a more reusable way.
Saving a user from defining multiple variations of the same task.

These two tasks can now be combined into a \ensuremath{\Conid{Circuit}}.
This \ensuremath{\Conid{Circuit}} can be interpreted as inputting a list of \texttt{.lhs} files, which are sequentially compiled to \texttt{.tex} files by the \ensuremath{\Varid{mapC}} operator.
These files are \textit{then} built by the \ensuremath{\Varid{buildTexTask}} to produce a PDF file as output.

\noindent\begin{minipage}{\linewidth}
\begin{hscode}\SaveRestoreHook
\column{B}{@{}>{\hspre}l<{\hspost}@{}}%
\column{3}{@{}>{\hspre}l<{\hspost}@{}}%
\column{E}{@{}>{\hspre}l<{\hspost}@{}}%
\>[B]{}\Varid{buildPipeline}\mathbin{::}\Conid{String}\to \Conid{String}\to \Conid{Circuit}\codeskip '[\mskip1.5mu \Conid{Var}\mskip1.5mu]\codeskip '[\mskip1.5mu [\mskip1.5mu \Conid{String}\mskip1.5mu]\mskip1.5mu]\codeskip '[\mskip1.5mu \Conid{Var}\mskip1.5mu]\codeskip '[\mskip1.5mu \Conid{String}\mskip1.5mu]\codeskip \Conid{N1}{}\<[E]%
\\
\>[B]{}\Varid{buildPipeline}\codeskip \Varid{outputFileName}\codeskip \Varid{mainFileName}\mathrel{=}{}\<[E]%
\\
\>[B]{}\hsindent{3}{}\<[3]%
\>[3]{}\Varid{mapC}\codeskip \Varid{lhs2TexTask}{}\<[E]%
\\
\>[B]{}\hsindent{3}{}\<[3]%
\>[3]{}<\!\!\!\!-\!\!\!\!>{}\<[E]%
\\
\>[B]{}\hsindent{3}{}\<[3]%
\>[3]{}\Varid{buildTexTask}\codeskip \Varid{outputFileName}\codeskip \Varid{mainFileName}{}\<[E]%
\ColumnHook
\end{hscode}\resethooks
\end{minipage}

\subsection{Using the Circuit}
To use this \ensuremath{\Conid{Circuit}}, a \ensuremath{\Conid{Config}} data type is used to store the information needed within the system to build the project:

\begin{hscode}\SaveRestoreHook
\column{B}{@{}>{\hspre}l<{\hspost}@{}}%
\column{3}{@{}>{\hspre}l<{\hspost}@{}}%
\column{6}{@{}>{\hspre}l<{\hspost}@{}}%
\column{18}{@{}>{\hspre}c<{\hspost}@{}}%
\column{18E}{@{}l@{}}%
\column{22}{@{}>{\hspre}l<{\hspost}@{}}%
\column{E}{@{}>{\hspre}l<{\hspost}@{}}%
\>[B]{}\mathbf{data}\codeskip \Conid{Config}\mathrel{=}\Conid{Config}{}\<[E]%
\\
\>[B]{}\hsindent{3}{}\<[3]%
\>[3]{}\{\mskip1.5mu {}\<[6]%
\>[6]{}\Varid{mainFile}{}\<[18]%
\>[18]{}\mathbin{::}{}\<[18E]%
\>[22]{}\Conid{FilePath}{}\<[E]%
\\
\>[B]{}\hsindent{3}{}\<[3]%
\>[3]{},{}\<[6]%
\>[6]{}\Varid{outputName}{}\<[18]%
\>[18]{}\mathbin{::}{}\<[18E]%
\>[22]{}\Conid{String}{}\<[E]%
\\
\>[B]{}\hsindent{3}{}\<[3]%
\>[3]{},{}\<[6]%
\>[6]{}\Varid{lhsFiles}{}\<[18]%
\>[18]{}\mathbin{::}{}\<[18E]%
\>[22]{}[\mskip1.5mu \Conid{FilePath}\mskip1.5mu]{}\<[E]%
\\
\>[B]{}\hsindent{3}{}\<[3]%
\>[3]{}\mskip1.5mu\}{}\<[E]%
\\
\>[B]{}\hsindent{3}{}\<[3]%
\>[3]{}\mathbf{deriving}\codeskip (\Conid{Generic},\Conid{FromJSON},\Conid{Show}){}\<[E]%
\ColumnHook
\end{hscode}\resethooks

This data type uses record syntax to have name fields:
\begin{itemize}
  \item \ensuremath{\Varid{mainFile}} is the name of the root file that should be used for compilation.
  \item \ensuremath{\Varid{outputName}} is the desired name for the output PDF file.
  \item \ensuremath{\Varid{lhsFiles}} are all the literate haskell files required to build the \LaTeX document.
\end{itemize}

The \ensuremath{\Conid{Config}} data type also derives the \ensuremath{\Conid{Generic}} and \ensuremath{\Conid{FromJSON}} instance.
This allows it to be used in conjunction with a YAML file to specify these parameters.
The config can be loaded with:

\begin{hscode}\SaveRestoreHook
\column{B}{@{}>{\hspre}l<{\hspost}@{}}%
\column{E}{@{}>{\hspre}l<{\hspost}@{}}%
\>[B]{}\Varid{loadConfig}\mathbin{::}\Conid{IO}\codeskip \Conid{Config}{}\<[E]%
\\
\>[B]{}\Varid{loadConfig}\mathrel{=}\Varid{loadYamlSettings}\codeskip [\mskip1.5mu \text{\ttfamily \char34 dissertation.tex-build\char34}\mskip1.5mu]\codeskip [\mskip1.5mu \mskip1.5mu]\codeskip \Varid{ignoreEnv}{}\<[E]%
\ColumnHook
\end{hscode}\resethooks

An example config file can be seen in Figure~\ref{fig:examples-lhs2tex-config-file}.

\begin{figure}[ht]
  \centering
  \begin{lstlisting}
  mainFile: dissertation.lhs
  outputName: dissertation
  lhsFiles: [dissertation.lhs
          , chapters/introduction.lhs
          , chapters/background.lhs
          , chapters/the-language.lhs
          , chapters/implementation.lhs
          , chapters/evaluation.lhs
          , chapters/examples.lhs]
  \end{lstlisting}
  \caption{An example config file for the lhs2TeX build system}
  \label{fig:examples-lhs2tex-config-file}
\end{figure}

To be able to use the system a \ensuremath{\Varid{main}} function is defined, which will serve as the entry point to the executable:

\noindent\begin{minipage}{\linewidth}
\begin{hscode}\SaveRestoreHook
\column{B}{@{}>{\hspre}l<{\hspost}@{}}%
\column{3}{@{}>{\hspre}l<{\hspost}@{}}%
\column{7}{@{}>{\hspre}l<{\hspost}@{}}%
\column{10}{@{}>{\hspre}l<{\hspost}@{}}%
\column{E}{@{}>{\hspre}l<{\hspost}@{}}%
\>[B]{}\Varid{main}\mathbin{::}\Conid{IO}\codeskip (){}\<[E]%
\\
\>[B]{}\Varid{main}\mathrel{=}\mathbf{do}{}\<[E]%
\\
\>[B]{}\hsindent{3}{}\<[3]%
\>[3]{}\Varid{config}\leftarrow \Varid{loadConfig}{}\<[E]%
\\
\>[B]{}\hsindent{3}{}\<[3]%
\>[3]{}\Varid{n}{}\<[10]%
\>[10]{}\leftarrow \Varid{startNetwork}\codeskip (\Varid{buildPipeline}\codeskip (\Varid{outputName}\codeskip \Varid{config})\codeskip (\Varid{mainFile}\codeskip \Varid{config}))\mathbin{::}\Conid{IO}{}\<[E]%
\\
\>[3]{}\hsindent{4}{}\<[7]%
\>[7]{}(\Conid{BasicNetwork}\codeskip '[\mskip1.5mu \Conid{Var}\mskip1.5mu]\codeskip '[\mskip1.5mu [\mskip1.5mu \Conid{String}\mskip1.5mu]\mskip1.5mu]\codeskip '[\mskip1.5mu \Conid{Var}\mskip1.5mu]\codeskip '[\mskip1.5mu \Conid{String}\mskip1.5mu]){}\<[E]%
\\[\blanklineskip]%
\>[B]{}\hsindent{3}{}\<[3]%
\>[3]{}\Varid{inputJobUUID}\leftarrow \Varid{genJobUUID}{}\<[E]%
\\
\>[B]{}\hsindent{3}{}\<[3]%
\>[3]{}\Varid{inputTaskUUID}\leftarrow \Varid{genTaskUUID}{}\<[E]%
\\
\>[B]{}\hsindent{3}{}\<[3]%
\>[3]{}(\Varid{inputVar}\mathbin{::}\Conid{Var}\codeskip [\mskip1.5mu \Conid{FilePath}\mskip1.5mu])\leftarrow \Varid{empty}\codeskip \Varid{inputTaskUUID}\codeskip \Varid{inputJobUUID}{}\<[E]%
\\
\>[B]{}\hsindent{3}{}\<[3]%
\>[3]{}\Varid{save}\codeskip \Varid{inputVar}\codeskip (\Varid{lhsFiles}\codeskip \Varid{config}){}\<[E]%
\\
\>[B]{}\hsindent{3}{}\<[3]%
\>[3]{}\Varid{write}\codeskip \Varid{inputJobUUID}\codeskip (\Conid{HCons'}\codeskip \Varid{inputVar}\codeskip \Conid{HNil'})\codeskip \Varid{n}{}\<[E]%
\\
\>[B]{}\hsindent{3}{}\<[3]%
\>[3]{}\anonymous \leftarrow \Varid{read}\codeskip \Varid{n}{}\<[E]%
\\
\>[B]{}\hsindent{3}{}\<[3]%
\>[3]{}\Varid{stopNetwork}\codeskip \Varid{n}{}\<[E]%
\ColumnHook
\end{hscode}\resethooks
\end{minipage}

The \ensuremath{\Varid{main}} function in the build system has 5 steps:

\begin{enumerate}
  \item The config file is loaded.
  \item A network is started based on the \ensuremath{\Varid{buildDiss}} circuit.
        The type of network to be started has to be annotated, so that the type system knows which \ensuremath{\Conid{Network}} instance to use.
  \item The build job is input into the network, with the input values being a list of \texttt{.lhs} files to compile.
  \item A call is made to the blocking function \ensuremath{\Varid{read}}, although the outputs are not needed, the call to \ensuremath{\Varid{read}} is.
        This prevents the program ending before the network has complete processing values.
  \item Finally, the network is destroyed.
\end{enumerate}

The thesis this paper is based on makes use of this build system to be able include literate Haskell files.

\section{Monoidal Resource Theories}\label{app:resource-theories}

Resource theories~\cite{Coecke_2016} are a branch of mathematics that allow for the reasoning of questions surrounding resources, for example:
\textit{If I have some resources, can I make something?}
\textit{If I have some resources, how can I get what I want?}
Questions that are eerily familiar to the questions addressed by dataflow:
\textit{If I have some inputs, can I make an output?}
\textit{If I have some inputs, how can I get the output I want?}
Resource theories provide a way to answer these questions, making them an excellent source of inspiration for our dataflow language. They combine \textit{preorders} and \textit{symmetric monoids} together to describe collections of resources (or inputs) and dependencies between them.

\textbf{Symmetric Monoids} are monoids with the following additional axiom: $\forall x, y \in X$, $x \otimes y = y \otimes x$ (Symmetry).

\textbf{\acp{SMP}} A symmetric monoidal structure $(X, I, \otimes)$ on a preorder $(X, \le)$ encapsulates this respect of the preorder as an additional axiom:
$\forall x_1, x_2, y_1, y_2 \in X$, if $x_1 \le y_1$ and $x_2 \le y_2$, then $x_1 \otimes x_2 \le y_1 \otimes y_2$ (Monotonicity).
One example is where $X$ is a collection of resources, $\otimes$ combines resources together, and $\le$ defines dependencies between resources.

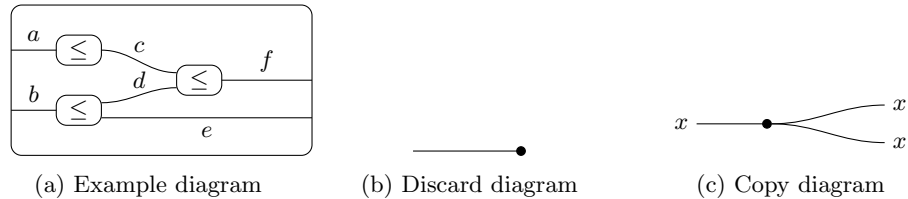
\begin{figure}[ht]

  \begin{subfigure}[b]{0.3\textwidth}
    \centering
    \begin{tikzpicture}[node distance={15mm}, main/.style = {draw, thick}, scale=0.4]

      \draw[rounded corners] (0, 0) rectangle (10, 5) {};
      \draw[rounded corners] (1.5, 1) rectangle (3, 2) node[pos=0.5] {$\le$};
      \draw[rounded corners] (1.5, 4) rectangle (3, 3) node[pos=0.5] {$\le$};
      \draw[rounded corners] (5.5, 2) rectangle (7, 3) node[pos=0.5] {$\le$};

      \draw (0, 3.5) -- (1.5, 3.5) node [midway, above] {$a$};
      \draw (0, 1.5) -- (1.5, 1.5) node [midway, above] {$b$};
      \draw (3, 1.25) -- (10, 1.25) node [midway, below] {$e$};

      \draw (3, 3.5) edge[out=0,in=180] node[yshift=2mm] {$c$} (5.5, 2.75);
      \draw (3, 1.75) edge[out=0,in=180] node[yshift=2mm] {$d$} (5.5, 2.25);

      \draw (7, 2.5) -- (10, 2.5) node [midway, above] {$f$};

    \end{tikzpicture}

    \caption{Example diagram}
    \label{fig:bg-wiring-diagram-example}
  \end{subfigure}
  \hfill
  \begin{subfigure}[b]{0.3\textwidth}
    \centering
    \begin{tikzpicture}
      \draw[-{Circle[black]}] (0,0) -- (1.5, 0);
    \end{tikzpicture}
    \caption{Discard diagram}
    \label{fig:discard-wiring-diagram}
  \end{subfigure}
  \hfill
  \begin{subfigure}[b]{0.3\textwidth}
    \centering
      \begin{tikzpicture}

      \node at (-0.2,0) {$x$};
      \draw[-{Circle[black]}] (0,0) -- (1, 0);

      \draw (1, 0) edge[out=0,in=180] (2.5, 0.25);
      \draw (1, 0) edge[out=0,in=180] (2.5, -0.25);

      \node at (2.7,0.25) {$x$};
      \node at (2.7,-0.25) {$x$};
    \end{tikzpicture}
    \caption{Copy diagram}
    \label{fig:copy-wiring-diagram}
  \end{subfigure}

\end{figure}

\textbf{Wiring Diagrams:}
A graphical representation of \acp{SMP} is a wiring diagram, similar to \acp{DAG}. A wiring diagram is made up of boxes that can have multiple inputs and outputs.
The boxes can be arranged in series or in parallel.
Figure~\ref{fig:bg-wiring-diagram-example}, shows an example wiring diagram.

A wiring diagram formalises a \ac{SMP}, with each element $x \in X$ existing as the label on a wire.
Two wires, $x$ and $y$, drawn in parallel are considered to be the monoidal product $x \otimes y$.
The monodial unit is defined as a wire with the label $I$ or no wire.

\begin{center}
\begin{tikzpicture}
\draw (0, 0) -- (1.5, 0) node[midway, below] {$y$};
\draw (0, 0.25) -- (1.5, 0.25) node[midway, above] {$x$};
\end{tikzpicture}
\end{center}

A box connects parallel wires on the left to parallel wires on the right.
A wiring diagram is considered valid if the monoidal product of the left is less than the right.

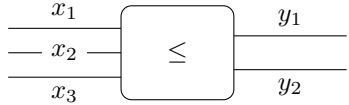
\begin{center}
\begin{tikzpicture}
\draw (0, -0.2) -- (1.5, -0.2) node[midway, below] {$x_3$};
\draw (0, 0.125) -- (1.5, 0.125) node[midway, fill=white] {$x_2$};
\draw (0, 0.45) -- (1.5, 0.45) node[midway, above] {$x_1$};
\draw[rounded corners] (1.5, -0.5) rectangle (3, 0.75) node[pos=0.5] {$\le$};
\draw (3, -0.1) -- (4.5, -0.1) node[midway, below] {$y_2$};
\draw (3, 0.35) -- (4.5, 0.35) node[midway, above] {$y_1$};
\end{tikzpicture}
\end{center}

This example wiring diagram corresponds to the inequality $x_1 \otimes x_2 \otimes x_3 \le y_1 \otimes y_2$, which corresponds to the idea that $x_1$, $x_2$, and $x_3$ are required to get $y_1$, and $y_2$.

Each axiom in a \ac{SMP} has a corresponding graphical form, using wiring diagrams.

\paragraph{Reflexivity}
The reflexivity law states that $x \le x$, this states that a diagram of one wire is valid.

\begin{center}
\begin{tikzpicture}
\draw (0, 0) -- (2, 0) node[midway, below] {$x$};
\end{tikzpicture}
\end{center}

This law corresponds to the idea that a resource is preserved.

\paragraph{Transitivity}
The transitivity law says that if $x \le y$ and $y \le z$ then $x \le z$. This corresponds to connecting two diagrams together in sequence.
If both of the diagrams

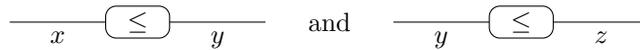
\begin{center}
\begin{tikzpicture}[scale=0.85]
\draw (0, 0) -- (1.5, 0) node[midway, below] {$x$};
\draw[rounded corners] (1.5, -0.25) rectangle (2.5, 0.25) node[pos=0.5] {$\le$};
\draw (2.5, 0) -- (4, 0) node[midway, below] {$y$};

\node at (5, 0) {and};

\draw (6, 0) -- (7.5, 0) node[midway, below] {$y$};
\draw[rounded corners] (7.5, -0.25) rectangle (8.5, 0.25) node[pos=0.5] {$\le$};
\draw (8.5, 0) -- (10, 0) node[midway, below] {$z$};
\end{tikzpicture}
\end{center}

\noindent
are valid, then they can be joined together to obtain another valid diagram.

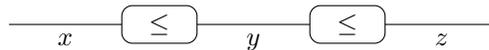
\begin{center}
\begin{tikzpicture}
\draw (0, 0) -- (1.5, 0) node[midway, below] {$x$};
\draw[rounded corners] (1.5, -0.25) rectangle (2.5, 0.25) node[pos=0.5] {$\le$};
\draw (2.5, 0) -- (4, 0) node[midway, below] {$y$};
\draw[rounded corners] (4, -0.25) rectangle (5, 0.25) node[pos=0.5] {$\le$};
\draw (5, 0) -- (6.5, 0) node[midway, below] {$z$};
\end{tikzpicture}
\end{center}

If a box is considered a task that can transform values, then this law corresponds to the idea that two tasks can be composed in sequence, with the output of one being the input to the next.

\paragraph{Unitality}
The unitality law states that $I \otimes x = x$ and $x \otimes I = x$, this means that a blank space can be ignored and that diagrams such as

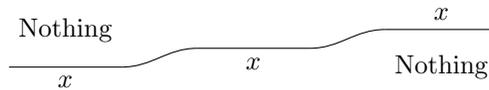
\begin{center}
\begin{tikzpicture}
\draw (0, 0) -- (1.5, 0) node[midway, below] {$x$};
\node at (0.75, 0.5) {Nothing};

\draw (1.5, 0) edge[out=0,in=180] (2.5, 0.25);

\draw (2.5, 0.25) -- (4, 0.25) node[midway, below] {$x$};

\draw (4, 0.25) edge[out=0,in=180] (5, 0.5);

\node at (5.75, 0) {Nothing};
\draw (5, 0.5) -- (6.5, 0.5) node[midway, above] {$x$};
\end{tikzpicture}
\end{center}

\noindent
are valid.

\paragraph{Associativity}
The associativity law says that $(x \otimes y) \otimes z = x \otimes (y \otimes z)$, this states that diagrams can be built from either the top or bottom.
This means that the order of grouping resources does not matter.
In reality it is trivial to see how this is true with wires:

\begin{center}
\begin{tikzpicture}
\draw (0, -0.4) -- (1.5, -0.4) node[midway, below] {$z$};
\draw (0, 0.125) -- (1.5, 0.125) node[midway, below] {$y$};
\draw (0, 0.45) -- (1.5, 0.45) node[midway, above] {$x$};

\node at (2.5, 0.125) {$=$};

\draw (3.5, -0.4) -- (5, -0.4) node[midway, below] {$z$};
\draw (3.5, -0.075) -- (5, -0.075) node[midway, above] {$y$};
\draw (3.5, 0.45) -- (5, 0.45) node[midway, above] {$x$};
\end{tikzpicture}
\end{center}

\paragraph{Symmetry}
The symmetry law states that $x\ \otimes\ y = y\ \otimes\ x$, this encodes the notion that a diagram is still valid even if the wires cross.

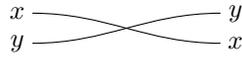
\begin{center}
\begin{tikzpicture}
\node at (-0.2,0.4) {$x$};
\node at (-0.2,0) {$y$};
\draw (0, 0) edge[out=0,in=180] (2.5, 0.4) ;
\draw (0, 0.4) edge[out=0,in=180] (2.5, 0);
\node at (2.7,0.4) {$y$};
\node at (2.7,0) {$x$};
\end{tikzpicture}
\end{center}

\paragraph{Monotonicity}
Monotonicity states that, if $x_1 \le y_1$ and $x_2 \le y_2$, then $x_1 \otimes x_2 \le y_1 \otimes y_2$. This can be thought of as stacking two boxes on top of each other:

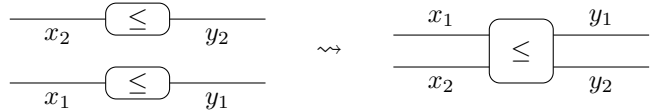
\begin{center}
\begin{tikzpicture}[scale = 0.85]
\draw (0, 0) -- (1.5, 0) node[midway, below] {$x_1$};
\draw[rounded corners] (1.5, -0.25) rectangle (2.5, 0.25) node[pos=0.5] {$\le$};
\draw (2.5, 0) -- (4, 0) node[midway, below] {$y_1$};

\draw (0, 1) -- (1.5, 1) node[midway, below] {$x_2$};
\draw[rounded corners] (1.5, 0.75) rectangle (2.5, 1.25) node[pos=0.5] {$\le$};
\draw (2.5, 1) -- (4, 1) node[midway, below] {$y_2$};

\node at (5, 0.5) {$\leadsto$};

\draw (6, 0.25) -- (7.5, 0.25) node[midway, below] {$x_2$};
\draw (6, 0.75) -- (7.5, 0.75) node[midway, above] {$x_1$};
\draw[rounded corners] (7.5, 0) rectangle (8.5, 1) node[pos=0.5] {$\le$};
\draw (8.5, 0.25) -- (10, 0.25) node[midway, below] {$y_2$};
\draw (8.5, 0.75) -- (10, 0.75) node[midway, above] {$y_1$};
\end{tikzpicture}
\end{center}

This law conceptualises the idea that when resources are combined, the dependencies are respected.

Resource theories can level up to include additional axioms that encode desirable things to do with data, such as deleting it, or duplicating it.

\textit{Discard Axiom}
There are times when there is no longer need to keep a value, it would be beneficial if it could be discarded.
In a wiring diagram this is represented as a wire that ends (Figure \ref{fig:discard-wiring-diagram}).
It corresponds to the idea that resources can be destroyed when they are no longer needed.

\textit{Copy Axiom}
The final axiom to add is the notion of copying a value: $\forall x \in X, x \le x + x$.
This can be represented in wiring diagram as a split wire (Figure \ref{fig:copy-wiring-diagram}).
This embodies the idea that it is possible to duplicate a resource.

\section{Circuit Forms a Symmetric Monoidal Preorder}\label{app:SMP-proof}

For simplicity only the \ensuremath{\Varid{insApplied}} and \ensuremath{\Varid{outsApplied}} type parameters will be used to formalise a \ensuremath{\Conid{Circuit}} --- all other type parameters are only required to aid GHC in compilation.

A preorder is defined over tasks and \ensuremath{\Conid{DataStore}}s.
The preorder relation $\le$, can be used to describe the dependencies in the \ensuremath{\Conid{DataStore}}s, with a task being able to transform \ensuremath{\Conid{DataStore}}s into new \ensuremath{\Conid{DataStore}}s.
The relation is defined over the set $X$, which describes the set of all possible \ensuremath{\Conid{DataStore}}s.

The monoidal product $\otimes$ can be thought of as the concatenation of multiple \ensuremath{\Conid{DataStore}}s into type-lists.
For example the monoidal product of $(f\:a) \otimes (g\:b) = $ \ensuremath{'[\mskip1.5mu \Varid{f}\codeskip \Varid{a}\mskip1.5mu]:\!\!+\!\!+'[\mskip1.5mu \Varid{g}\codeskip \Varid{b}\mskip1.5mu]\mathord{\sim}'[\mskip1.5mu \Varid{f}\codeskip \Varid{a},\Varid{g}\codeskip \Varid{b}\mskip1.5mu]}.
The monoidal unit, is tricky to define as it has no real meaning within a \ensuremath{\Conid{Circuit}}, however it could be considered the empty \ensuremath{\Conid{DataStore}}: \ensuremath{'[\mskip1.5mu \mskip1.5mu]}.

The axioms are then satisfied as follows:
\begin{enumerate}
  \item Reflexivity --- this is the \ensuremath{\Varid{id}\mathbin{::}\Conid{Circuit}\codeskip '[\mskip1.5mu \Varid{f}\codeskip \Varid{a}\mskip1.5mu]\codeskip '[\mskip1.5mu \Varid{f}\codeskip \Varid{a}\mskip1.5mu]} constructor, it represents a straight line with the same input and output.
  \item Transitivity --- this is the \ensuremath{<\!\!\!\!-\!\!\!\!>\mathbin{::}\Conid{Circuit}\codeskip \Varid{x}\codeskip \Varid{y}\to \Conid{Circuit}\codeskip \Varid{y}\codeskip \Varid{z}\to \Conid{Circuit}\codeskip \Varid{x}\codeskip \Varid{z}} constructor, it allows for circuits to be placed in sequence.
  \item Monotonicity --- this is the \ensuremath{\mathbin{<>}\mathbin{::}\Conid{Circuit}\codeskip \Varid{x}_{1}\codeskip \Varid{y}_{1}\to \Conid{Circuit}\codeskip \Varid{x}_{2}\codeskip \Varid{y}_{2}\to \Conid{Circuit}\codeskip (\Varid{x}_{1}:\!\!+\!\!+\Varid{x}_{2})\codeskip (\Varid{y}_{1}:\!\!+\!\!+\Varid{y}_{2})} constructor. This can place circuits next to each other.
  \item Unitality --- given the monoidal unit \ensuremath{'[\mskip1.5mu \mskip1.5mu]} and a \ensuremath{\Conid{DataStore}} \ensuremath{\Varid{xs}}, then the rules hold true: \ensuremath{'[\mskip1.5mu \mskip1.5mu]:\!\!+\!\!+\Varid{xs}\mathord{\sim}\Varid{xs}} and \ensuremath{\Varid{xs}:\!\!+\!\!+'[\mskip1.5mu \mskip1.5mu]\mathord{\sim}\Varid{xs}}.
  \item Associativity --- given three \ensuremath{\Conid{DataStore}}s: \ensuremath{\Varid{xs}}, \ensuremath{\Varid{ys}}, \ensuremath{\Varid{zs}}. Since concatenation of lists is associative then this rule holds: \ensuremath{(\Varid{xs}:\!\!+\!\!+\Varid{ys}):\!\!+\!\!+\Varid{zs}\mathord{\sim}\Varid{xs}:\!\!+\!\!+(\Varid{ys}:\!\!+\!\!+\Varid{zs})}.
  \item Symmetry --- this is the \ensuremath{\Varid{swap}\mathbin{::}\Conid{Circuit}\codeskip '[\mskip1.5mu \Varid{f}\codeskip \Varid{a},\Varid{g}\codeskip \Varid{b}\mskip1.5mu]\codeskip '[\mskip1.5mu \Varid{g}\codeskip \Varid{b},\Varid{f}\codeskip \Varid{a}\mskip1.5mu]} constructor, it allows for values to swap over.
  \item Delete Axiom --- this is satisfied by the \ensuremath{\Varid{dropL}\mathbin{::}\Conid{Circuit}\codeskip '[\mskip1.5mu \Varid{f}\codeskip \Varid{a},\Varid{g}\codeskip \Varid{b}\mskip1.5mu]\codeskip '[\mskip1.5mu \Varid{g}\codeskip \Varid{b}\mskip1.5mu]} and \ensuremath{\Varid{dropR}\mathbin{::}\Conid{Circuit}\codeskip '[\mskip1.5mu \Varid{f}\codeskip \Varid{a},\Varid{g}\codeskip \Varid{b}\mskip1.5mu]\codeskip '[\mskip1.5mu \Varid{f}\codeskip \Varid{a}\mskip1.5mu]}.
    Although this does not directly fit with the axiom, it also has to ensure the constraint on a circuit that there must always be 1 output value.
  \item Copy Axiom --- this is the \ensuremath{\Varid{replicate}\mathbin{::}\Conid{Circuit}\codeskip '[\mskip1.5mu \Varid{f}\codeskip \Varid{a}\mskip1.5mu]\codeskip '[\mskip1.5mu \Varid{f}\codeskip \Varid{a},\Varid{f}\codeskip \Varid{a}\mskip1.5mu]} constructor. It allows for a \ensuremath{\Conid{DataStore}} to be duplicated.
\end{enumerate}

By satisfying all the axioms a \ensuremath{\Conid{Circuit}} is a \ac{SMP}.

\section{Beside Algebra}\label{sec:beside}
An instance of the algebra is defined as:
\begin{hscode}\SaveRestoreHook
\column{B}{@{}>{\hspre}l<{\hspost}@{}}%
\column{3}{@{}>{\hspre}l<{\hspost}@{}}%
\column{5}{@{}>{\hspre}l<{\hspost}@{}}%
\column{E}{@{}>{\hspre}l<{\hspost}@{}}%
\>[3]{}\mathbf{instance}\codeskip \Conid{BuildNetworkAlg}\codeskip \Conid{BasicNetwork}\codeskip \Conid{Beside}\codeskip \mathbf{where}{}\<[E]%
\\
\>[3]{}\hsindent{2}{}\<[5]%
\>[5]{}\Varid{buildNetworkAlg}\mathrel{=}\Varid{beside}{}\<[E]%
\ColumnHook
\end{hscode}\resethooks
This requires a \ensuremath{\Varid{beside}} function, however to define this function some extra tools are required.
The first is \ensuremath{\Varid{takeP}}, which will take the first \ensuremath{\Varid{n}} elements from a \ensuremath{\Conid{PipeList}}:
\begin{hscode}\SaveRestoreHook
\column{B}{@{}>{\hspre}l<{\hspost}@{}}%
\column{3}{@{}>{\hspre}l<{\hspost}@{}}%
\column{10}{@{}>{\hspre}l<{\hspost}@{}}%
\column{21}{@{}>{\hspre}l<{\hspost}@{}}%
\column{38}{@{}>{\hspre}c<{\hspost}@{}}%
\column{38E}{@{}l@{}}%
\column{41}{@{}>{\hspre}l<{\hspost}@{}}%
\column{E}{@{}>{\hspre}l<{\hspost}@{}}%
\>[3]{}\Varid{takeP}{}\<[10]%
\>[10]{}\mathbin{::}\Conid{SNat}\codeskip \Varid{n}\to \Conid{PipeList}\codeskip \Varid{fs}\codeskip \Varid{as}\to \Conid{PipeList}\codeskip (\Conid{Take}\codeskip \Varid{n}\codeskip \Varid{fs})\codeskip (\Conid{Take}\codeskip \Varid{n}\codeskip \Varid{as}){}\<[E]%
\\
\>[3]{}\Varid{takeP}\codeskip {}\<[10]%
\>[10]{}\Conid{SZero}\codeskip {}\<[21]%
\>[21]{}\anonymous {}\<[38]%
\>[38]{}\mathrel{=}{}\<[38E]%
\>[41]{}\Conid{PipeNil}{}\<[E]%
\\
\>[3]{}\Varid{takeP}\codeskip {}\<[10]%
\>[10]{}(\Conid{SSucc}\codeskip \anonymous )\codeskip {}\<[21]%
\>[21]{}\Conid{PipeNil}{}\<[38]%
\>[38]{}\mathrel{=}{}\<[38E]%
\>[41]{}\Conid{PipeNil}{}\<[E]%
\\
\>[3]{}\Varid{takeP}\codeskip {}\<[10]%
\>[10]{}(\Conid{SSucc}\codeskip \Varid{n})\codeskip {}\<[21]%
\>[21]{}(\Conid{PipeCons}\codeskip \Varid{x}\codeskip \Varid{xs}){}\<[38]%
\>[38]{}\mathrel{=}{}\<[38E]%
\>[41]{}\Conid{PipeCons}\codeskip \Varid{x}\codeskip (\Varid{takeP}\codeskip \Varid{n}\codeskip \Varid{xs}){}\<[E]%
\ColumnHook
\end{hscode}\resethooks
This makes use of the \ensuremath{\Conid{Take}} type family to take \ensuremath{\Varid{n}} elements from each of the type lists: \ensuremath{\Varid{fs}}, \ensuremath{\Varid{as}}, and \ensuremath{\Varid{xs}}.
It follows the same structure as the \ensuremath{\Varid{take}\mathbin{::}\Conid{Int}\to [\mskip1.5mu \Varid{a}\mskip1.5mu]\to [\mskip1.5mu \Varid{a}\mskip1.5mu]} defined in the \ensuremath{\Conid{Prelude}}.

The next function is \ensuremath{\Varid{dropP}}, it drops \ensuremath{\Varid{n}} elements from a \ensuremath{\Conid{PipeList}}:
\begin{hscode}\SaveRestoreHook
\column{B}{@{}>{\hspre}l<{\hspost}@{}}%
\column{3}{@{}>{\hspre}l<{\hspost}@{}}%
\column{10}{@{}>{\hspre}l<{\hspost}@{}}%
\column{21}{@{}>{\hspre}l<{\hspost}@{}}%
\column{38}{@{}>{\hspre}l<{\hspost}@{}}%
\column{E}{@{}>{\hspre}l<{\hspost}@{}}%
\>[3]{}\Varid{dropP}{}\<[10]%
\>[10]{}\mathbin{::}\Conid{SNat}\codeskip \Varid{n}\to \Conid{PipeList}\codeskip \Varid{fs}\codeskip \Varid{as}\to \Conid{PipeList}\codeskip (\Conid{Drop}\codeskip \Varid{n}\codeskip \Varid{fs})\codeskip (\Conid{Drop}\codeskip \Varid{n}\codeskip \Varid{as}){}\<[E]%
\\
\>[3]{}\Varid{dropP}\codeskip {}\<[10]%
\>[10]{}\Conid{SZero}\codeskip {}\<[21]%
\>[21]{}\Varid{l}{}\<[38]%
\>[38]{}\mathrel{=}\Varid{l}{}\<[E]%
\\
\>[3]{}\Varid{dropP}\codeskip {}\<[10]%
\>[10]{}(\Conid{SSucc}\codeskip \anonymous )\codeskip {}\<[21]%
\>[21]{}\Conid{PipeNil}{}\<[38]%
\>[38]{}\mathrel{=}\Conid{PipeNil}{}\<[E]%
\\
\>[3]{}\Varid{dropP}\codeskip {}\<[10]%
\>[10]{}(\Conid{SSucc}\codeskip \Varid{n})\codeskip {}\<[21]%
\>[21]{}(\Conid{PipeCons}\codeskip \anonymous \codeskip \Varid{xs}){}\<[38]%
\>[38]{}\mathrel{=}\Varid{dropP}\codeskip \Varid{n}\codeskip \Varid{xs}{}\<[E]%
\ColumnHook
\end{hscode}\resethooks
This function again follows the same structure as \ensuremath{\Varid{drop}\mathbin{::}\Conid{Int}\to [\mskip1.5mu \Varid{a}\mskip1.5mu]\to [\mskip1.5mu \Varid{a}\mskip1.5mu]} defined in the \ensuremath{\Conid{Prelude}}.
Both \ensuremath{\Varid{takeP}} and \ensuremath{\Varid{dropP}} are used to split the outputs of a network after \ensuremath{\Varid{n}} elements.
This requires the knowledge of what \ensuremath{\Varid{n}} is at the value level, however \ensuremath{\Varid{n}} is only stored at the type level as the argument \ensuremath{\Varid{nins}}.
To be able to recover this value the \ensuremath{\Conid{IsNat}} type class is used.
The \ensuremath{\Varid{recoverNIns}} function is able to direct the \ensuremath{\Conid{IsNat}} type class to the correct type argument,
and produces an \ensuremath{\Conid{SNat}} with the same value as that stored in the type.
\begin{hscode}\SaveRestoreHook
\column{B}{@{}>{\hspre}l<{\hspost}@{}}%
\column{3}{@{}>{\hspre}l<{\hspost}@{}}%
\column{5}{@{}>{\hspre}c<{\hspost}@{}}%
\column{5E}{@{}l@{}}%
\column{9}{@{}>{\hspre}l<{\hspost}@{}}%
\column{E}{@{}>{\hspre}l<{\hspost}@{}}%
\>[3]{}\Varid{recoverNIns}\mathbin{::}(\Conid{Length}\codeskip \Varid{bsS}\mathord{\sim}\Conid{Length}\codeskip \Varid{bsT},\Varid{nins}\mathord{\sim}\Conid{Length}\codeskip \Varid{bsS},\Conid{IsNat}\codeskip \Varid{nins},\Conid{Network}\codeskip \Varid{n}){}\<[E]%
\\
\>[3]{}\hsindent{2}{}\<[5]%
\>[5]{}\Rightarrow {}\<[5E]%
\>[9]{}(\Conid{N}\codeskip \Varid{n}\codeskip \Varid{asS}\codeskip \Varid{asT})\codeskip \Varid{bsS}\codeskip \Varid{bsT}\codeskip \Varid{csS}\codeskip \Varid{csT}\codeskip (\Varid{nins}\mathbin{::}\Conid{Nat})\to \Conid{SNat}\codeskip (\Conid{Length}\codeskip \Varid{bsS}){}\<[E]%
\\
\>[3]{}\Varid{recoverNIns}\codeskip \anonymous \mathrel{=}\Varid{nat}{}\<[E]%
\ColumnHook
\end{hscode}\resethooks
After splitting a network and generating two new networks, the outputs will need to be joined together again: this will require the appending of two \ensuremath{\Conid{PipeLists}}.
To do this an \ensuremath{\Conid{AppendP}} type class is defined:
\begin{hscode}\SaveRestoreHook
\column{B}{@{}>{\hspre}l<{\hspost}@{}}%
\column{3}{@{}>{\hspre}l<{\hspost}@{}}%
\column{5}{@{}>{\hspre}l<{\hspost}@{}}%
\column{14}{@{}>{\hspre}l<{\hspost}@{}}%
\column{E}{@{}>{\hspre}l<{\hspost}@{}}%
\>[3]{}\mathbf{class}\codeskip \Conid{AppendP}\codeskip \Varid{fs}\codeskip \Varid{as}\codeskip \Varid{gs}\codeskip \Varid{bs}\codeskip \mathbf{where}{}\<[E]%
\\
\>[3]{}\hsindent{2}{}\<[5]%
\>[5]{}\Varid{appendP}{}\<[14]%
\>[14]{}\mathbin{::}\Conid{PipeList}\codeskip \Varid{fs}\codeskip \Varid{as}\to \Conid{PipeList}\codeskip \Varid{gs}\codeskip \Varid{bs}\to \Conid{PipeList}\codeskip (\Varid{fs}:\!\!+\!\!+\Varid{gs})\codeskip (\Varid{as}:\!\!+\!\!+\Varid{bs}){}\<[E]%
\ColumnHook
\end{hscode}\resethooks
This type class has one function \ensuremath{\Varid{appendP}}, it is able to append two \ensuremath{\Conid{PipeLists}} together.
It makes use of the  \ensuremath{:\!\!+\!\!+} type family to append the type lists together.
The instances for this type class are made up of two cases: the base case and a recursive case.
\begin{hscode}\SaveRestoreHook
\column{B}{@{}>{\hspre}l<{\hspost}@{}}%
\column{3}{@{}>{\hspre}l<{\hspost}@{}}%
\column{5}{@{}>{\hspre}l<{\hspost}@{}}%
\column{14}{@{}>{\hspre}l<{\hspost}@{}}%
\column{31}{@{}>{\hspre}l<{\hspost}@{}}%
\column{35}{@{}>{\hspre}l<{\hspost}@{}}%
\column{E}{@{}>{\hspre}l<{\hspost}@{}}%
\>[3]{}\mathbf{instance}\codeskip \Conid{AppendP}\codeskip '[\mskip1.5mu \mskip1.5mu]\codeskip '[\mskip1.5mu \mskip1.5mu]\codeskip \Varid{gs}\codeskip \Varid{bs}\codeskip \mathbf{where}{}\<[E]%
\\
\>[3]{}\hsindent{2}{}\<[5]%
\>[5]{}\Varid{appendP}\codeskip {}\<[14]%
\>[14]{}\Conid{PipeNil}\codeskip {}\<[31]%
\>[31]{}\Varid{ys}{}\<[35]%
\>[35]{}\mathrel{=}\Varid{ys}{}\<[E]%
\ColumnHook
\end{hscode}\resethooks
\begin{hscode}\SaveRestoreHook
\column{B}{@{}>{\hspre}l<{\hspost}@{}}%
\column{3}{@{}>{\hspre}l<{\hspost}@{}}%
\column{5}{@{}>{\hspre}l<{\hspost}@{}}%
\column{14}{@{}>{\hspre}l<{\hspost}@{}}%
\column{31}{@{}>{\hspre}l<{\hspost}@{}}%
\column{35}{@{}>{\hspre}l<{\hspost}@{}}%
\column{E}{@{}>{\hspre}l<{\hspost}@{}}%
\>[3]{}\mathbf{instance}\codeskip (\Conid{AppendP}\codeskip \Varid{fs}\codeskip \Varid{as}\codeskip \Varid{gs}\codeskip \Varid{bs})\Rightarrow \Conid{AppendP}\codeskip (\Varid{f}\codeskip \,'\!\!\mathbin{:}\codeskip \Varid{fs})\codeskip (\Varid{a}\codeskip \,'\!\!\mathbin{:}\codeskip \Varid{as})\codeskip \Varid{gs}\codeskip \Varid{bs}\codeskip \mathbf{where}{}\<[E]%
\\
\>[3]{}\hsindent{2}{}\<[5]%
\>[5]{}\Varid{appendP}\codeskip {}\<[14]%
\>[14]{}(\Conid{PipeCons}\codeskip \Varid{x}\codeskip \Varid{xs})\codeskip {}\<[31]%
\>[31]{}\Varid{ys}{}\<[35]%
\>[35]{}\mathrel{=}\Conid{PipeCons}\codeskip \Varid{x}\codeskip (\Varid{appendP}\codeskip \Varid{xs}\codeskip \Varid{ys}){}\<[E]%
\ColumnHook
\end{hscode}\resethooks
The base case corresponds to having an empty list on the left, with some other list on the right. Here the list on the right is returned.
The recursive case, simply takes 1 element from the left hand side and conses it onto the from of a recursive call, with the rest of the left hand side.

It is now possible to define the \ensuremath{\Varid{beside}} function. The result is calculated in 4 steps, with helper functions for each step:

\begin{enumerate}
  \item Get the number of inputs (\ensuremath{\Varid{nins}}) on the left hand side of the \ensuremath{\Conid{Beside}} constructor.
        This will give the information needed to split the inputted accumulated network \ensuremath{\Varid{n}}.
  \item Split the network into a left and right hand side.
        This will retain the same input type to the network, as there is no information on how to split that.
        Only the output \ensuremath{\Conid{PipeList}} will be split into two parts.
  \item Translate the both the left and right network.
        This will perform the recursive step and generate two new networks with the networks from the left and right added to the accumulated network \ensuremath{\Varid{n}}.
  \item Join the networks back together.
        Now that the left and right hand side of this layer has been added to the accumulated network,
        the two sides need to be joined back together to get a single network that can be returned.
\end{enumerate}
\begin{hscode}\SaveRestoreHook
\column{B}{@{}>{\hspre}l<{\hspost}@{}}%
\column{3}{@{}>{\hspre}l<{\hspost}@{}}%
\column{4}{@{}>{\hspre}l<{\hspost}@{}}%
\column{5}{@{}>{\hspre}l<{\hspost}@{}}%
\column{7}{@{}>{\hspre}l<{\hspost}@{}}%
\column{9}{@{}>{\hspre}l<{\hspost}@{}}%
\column{10}{@{}>{\hspre}l<{\hspost}@{}}%
\column{11}{@{}>{\hspre}l<{\hspost}@{}}%
\column{12}{@{}>{\hspre}l<{\hspost}@{}}%
\column{13}{@{}>{\hspre}l<{\hspost}@{}}%
\column{14}{@{}>{\hspre}l<{\hspost}@{}}%
\column{16}{@{}>{\hspre}l<{\hspost}@{}}%
\column{17}{@{}>{\hspre}l<{\hspost}@{}}%
\column{18}{@{}>{\hspre}l<{\hspost}@{}}%
\column{25}{@{}>{\hspre}l<{\hspost}@{}}%
\column{E}{@{}>{\hspre}l<{\hspost}@{}}%
\>[3]{}\Varid{beside}\mathbin{::}\forall \Varid{asS}\hsforall \codeskip \Varid{asT}\codeskip \Varid{bsS}\codeskip \Varid{bsT}\codeskip \Varid{csS}\codeskip \Varid{csT}\codeskip (\Varid{nbs}\mathbin{::}\Conid{Nat}){}\<[E]%
\\
\>[3]{}\hsindent{2}{}\<[5]%
\>[5]{}\hsdot{\cdot }{.}{}\<[9]%
\>[9]{}\Conid{Beside}\codeskip {}\<[17]%
\>[17]{}(\Conid{AccuN}\codeskip {}\<[25]%
\>[25]{}\Conid{BasicNetwork}\codeskip \Varid{asS}\codeskip \Varid{asT})\codeskip \Varid{bsS}\codeskip \Varid{bsT}\codeskip \Varid{csS}\codeskip \Varid{csT}\codeskip \Varid{nbs}{}\<[E]%
\\
\>[3]{}\hsindent{2}{}\<[5]%
\>[5]{}\to {}\<[9]%
\>[9]{}\Conid{IO}\codeskip {}\<[17]%
\>[17]{}((\Conid{AccuN}\codeskip \Conid{BasicNetwork}\codeskip \Varid{asS}\codeskip \Varid{asT})\codeskip \Varid{bsS}\codeskip \Varid{bsT}\codeskip \Varid{csS}\codeskip \Varid{csT}\codeskip \Varid{nbs}){}\<[E]%
\\
\>[3]{}\Varid{beside}\codeskip (\Conid{Beside}\codeskip \Varid{l}\codeskip \Varid{r})\mathrel{=}\keyw{return}\mathbin{\$}\Conid{AccuN}\codeskip (\lambda \Varid{n}\to \mathbf{do}{}\<[E]%
\\
\>[3]{}\hsindent{4}{}\<[7]%
\>[7]{}\mathbf{let}\codeskip \Varid{nins}\mathrel{=}\Varid{circuitIns}\codeskip \Varid{l}{}\<[E]%
\\
\>[3]{}\hsindent{4}{}\<[7]%
\>[7]{}(\Varid{nL}{}\<[12]%
\>[12]{},\Varid{nR}{}\<[18]%
\>[18]{})\leftarrow \Varid{splitNetwork}\codeskip \Varid{nins}\codeskip \Varid{n}{}\<[E]%
\\
\>[3]{}\hsindent{4}{}\<[7]%
\>[7]{}(\Varid{newL},\Varid{newR})\leftarrow \Varid{translate}\codeskip \Varid{nins}\codeskip (\Varid{nL},\Varid{nR})\codeskip (\Varid{l},\Varid{r}){}\<[E]%
\\
\>[3]{}\hsindent{4}{}\<[7]%
\>[7]{}\Varid{joinNetwork}\codeskip (\Varid{newL},\Varid{newR}){}\<[E]%
\\
\>[3]{}\hsindent{2}{}\<[5]%
\>[5]{}){}\<[E]%
\\
\>[3]{}\hsindent{1}{}\<[4]%
\>[4]{}\mathbf{where}{}\<[E]%
\\
\>[4]{}\hsindent{1}{}\<[5]%
\>[5]{}\Varid{splitNetwork}\mathbin{::}\Conid{SNat}\codeskip \Varid{nbsL}\to \Conid{BasicNetwork}\codeskip \Varid{asS}\codeskip \Varid{asT}\codeskip \Varid{bsS}\codeskip \Varid{bsT}{}\<[E]%
\\
\>[5]{}\hsindent{2}{}\<[7]%
\>[7]{}\to \Conid{IO}\codeskip ({}\<[16]%
\>[16]{}\Conid{BasicNetwork}\codeskip \Varid{asS}\codeskip \Varid{asT}\codeskip (\Conid{Take}\codeskip \Varid{nbsL}\codeskip \Varid{bsS})\codeskip (\Conid{Take}\codeskip \Varid{nbsL}\codeskip \Varid{bsT}),{}\<[E]%
\\
\>[16]{}\Conid{BasicNetwork}\codeskip \Varid{asS}\codeskip \Varid{asT}\codeskip (\Conid{Drop}\codeskip \Varid{nbsL}\codeskip \Varid{bsS})\codeskip (\Conid{Drop}\codeskip \Varid{nbsL}\codeskip \Varid{bsT})){}\<[E]%
\\
\>[4]{}\hsindent{1}{}\<[5]%
\>[5]{}\Varid{splitNetwork}\codeskip \Varid{nbs}\codeskip \Varid{n}\mathrel{=}\keyw{return}{}\<[E]%
\\
\>[5]{}\hsindent{2}{}\<[7]%
\>[7]{}({}\<[10]%
\>[10]{}\Conid{BasicNetwork}\codeskip (\Varid{threads}\codeskip \Varid{n})\codeskip (\Varid{ins}\codeskip \Varid{n})\codeskip (\Varid{takeP}\codeskip \Varid{nbs}\codeskip (\Varid{outs}\codeskip \Varid{n})),{}\<[E]%
\\
\>[10]{}\Conid{BasicNetwork}\codeskip (\Varid{threads}\codeskip \Varid{n})\codeskip (\Varid{ins}\codeskip \Varid{n})\codeskip (\Varid{dropP}\codeskip \Varid{nbs}\codeskip (\Varid{outs}\codeskip \Varid{n}))){}\<[E]%
\\[\blanklineskip]%
\>[4]{}\hsindent{1}{}\<[5]%
\>[5]{}\Varid{translate}\mathbin{::}\Conid{SNat}\codeskip \Varid{nbsL}{}\<[E]%
\\
\>[5]{}\hsindent{2}{}\<[7]%
\>[7]{}\to ({}\<[13]%
\>[13]{}\Conid{BasicNetwork}\codeskip \Varid{asS}\codeskip \Varid{asT}\codeskip (\Conid{Take}\codeskip \Varid{nbsL}\codeskip \Varid{bsS})\codeskip (\Conid{Take}\codeskip \Varid{nbsL}\codeskip \Varid{bsT}),{}\<[E]%
\\
\>[13]{}\Conid{BasicNetwork}\codeskip \Varid{asS}\codeskip \Varid{asT}\codeskip (\Conid{Drop}\codeskip \Varid{nbsL}\codeskip \Varid{bsS})\codeskip (\Conid{Drop}\codeskip \Varid{nbsL}\codeskip \Varid{bsT})){}\<[E]%
\\
\>[5]{}\hsindent{2}{}\<[7]%
\>[7]{}\to ({}\<[13]%
\>[13]{}(\Conid{AccuN}\codeskip \Conid{BasicNetwork}\codeskip \Varid{asS}\codeskip \Varid{asT})\codeskip (\Conid{Take}\codeskip \Varid{nbsL}\codeskip \Varid{bsS})\codeskip {}\<[E]%
\\
\>[13]{}\hsindent{3}{}\<[16]%
\>[16]{}(\Conid{Take}\codeskip \Varid{nbsL}\codeskip \Varid{bsT})\codeskip \Varid{csLS}\codeskip \Varid{csLT}\codeskip \Varid{nbsL},{}\<[E]%
\\
\>[13]{}(\Conid{AccuN}\codeskip \Conid{BasicNetwork}\codeskip \Varid{asS}\codeskip \Varid{asT})\codeskip (\Conid{Drop}\codeskip \Varid{nbsL}\codeskip \Varid{bsS})\codeskip {}\<[E]%
\\
\>[13]{}\hsindent{3}{}\<[16]%
\>[16]{}(\Conid{Drop}\codeskip \Varid{nbsL}\codeskip \Varid{bsT})\codeskip \Varid{csRS}\codeskip \Varid{csRT}\codeskip \Varid{nbsR}){}\<[E]%
\\
\>[5]{}\hsindent{2}{}\<[7]%
\>[7]{}\to \Conid{IO}\codeskip ({}\<[16]%
\>[16]{}\Conid{BasicNetwork}\codeskip \Varid{asS}\codeskip \Varid{asT}\codeskip \Varid{csLS}\codeskip \Varid{csLT},{}\<[E]%
\\
\>[16]{}\Conid{BasicNetwork}\codeskip \Varid{asS}\codeskip \Varid{asT}\codeskip \Varid{csRS}\codeskip \Varid{csRT}){}\<[E]%
\\
\>[4]{}\hsindent{1}{}\<[5]%
\>[5]{}\Varid{translate}\codeskip \anonymous \codeskip (\Varid{nL},\Varid{nR})\codeskip (\Conid{N}\codeskip \Varid{cL},\Conid{N}\codeskip \Varid{cR})\mathrel{=}\mathbf{do}{}\<[E]%
\\
\>[5]{}\hsindent{2}{}\<[7]%
\>[7]{}\Varid{nL'}\leftarrow \Varid{cL}\codeskip \Varid{nL}{}\<[E]%
\\
\>[5]{}\hsindent{2}{}\<[7]%
\>[7]{}\Varid{nR'}\leftarrow \Varid{cR}\codeskip \Varid{nR}{}\<[E]%
\\
\>[5]{}\hsindent{2}{}\<[7]%
\>[7]{}\keyw{return}\codeskip (\Varid{nL'},\Varid{nR'}){}\<[E]%
\\[\blanklineskip]%
\>[4]{}\hsindent{1}{}\<[5]%
\>[5]{}\Varid{joinNetwork}\mathbin{::}(\Conid{AppendP}\codeskip \Varid{csLS}\codeskip \Varid{csLT}\codeskip \Varid{csRS}\codeskip \Varid{csRT}){}\<[E]%
\\
\>[5]{}\hsindent{2}{}\<[7]%
\>[7]{}\Rightarrow {}\<[11]%
\>[11]{}({}\<[14]%
\>[14]{}\Conid{BasicNetwork}\codeskip \Varid{asS}\codeskip \Varid{asT}\codeskip \Varid{csLS}\codeskip \Varid{csLT}{}\<[E]%
\\
\>[11]{},{}\<[14]%
\>[14]{}\Conid{BasicNetwork}\codeskip \Varid{asS}\codeskip \Varid{asT}\codeskip \Varid{csRS}\codeskip \Varid{csRT}){}\<[E]%
\\
\>[5]{}\hsindent{2}{}\<[7]%
\>[7]{}\to \Conid{IO}\codeskip ({}\<[16]%
\>[16]{}\Conid{BasicNetwork}\codeskip \Varid{asS}\codeskip \Varid{asT}\codeskip (\Varid{csLS}:\!\!+\!\!+\Varid{csRS})\codeskip (\Varid{csLT}:\!\!+\!\!+\Varid{csRT})){}\<[E]%
\\
\>[4]{}\hsindent{1}{}\<[5]%
\>[5]{}\Varid{joinNetwork}\codeskip (\Varid{nL},\Varid{nR})\mathrel{=}\keyw{return}\mathbin{\$}\Conid{BasicNetwork}{}\<[E]%
\\
\>[5]{}\hsindent{6}{}\<[11]%
\>[11]{}(\Varid{nub}\codeskip (\Varid{threads}\codeskip \Varid{nL}\plus \Varid{threads}\codeskip \Varid{nR})){}\<[E]%
\\
\>[5]{}\hsindent{6}{}\<[11]%
\>[11]{}(\Varid{ins}\codeskip \Varid{nL}){}\<[E]%
\\
\>[5]{}\hsindent{6}{}\<[11]%
\>[11]{}(\Varid{outs}\codeskip \Varid{nL}\mathbin{\text{\`{}}\Varid{appendP}\text{\`{}}}\Varid{outs}\codeskip \Varid{nR}){}\<[E]%
\ColumnHook
\end{hscode}\resethooks
The \ensuremath{\Varid{splitNetwork}} function creates two new \ensuremath{\Conid{BasicNetwork}}s. To split the output values, \ensuremath{\Varid{takeP}}, and \ensuremath{\Varid{dropP}} are used.
\ensuremath{\Varid{translate}} performs the recursive step in the accumulating fold, which produces two new networks that include this layer.
\ensuremath{\Varid{joinNetwork}} takes the two new networks and appends the outputs with \ensuremath{\Varid{appendP}}.
It also has to append the thread ids from both sides, however, this will now include duplicates as threads were not split in \ensuremath{\Varid{splitNetwork}}.
To combat this \ensuremath{\Varid{nub}} is used, which returns a list containing all the unique values in the original.

Using the algebra in conjunction with \ensuremath{\Varid{icataM}_{5}},

\noindent\ensuremath{\Varid{buildBasicNetwork}} is now defined:
\begin{hscode}\SaveRestoreHook
\column{B}{@{}>{\hspre}l<{\hspost}@{}}%
\column{3}{@{}>{\hspre}l<{\hspost}@{}}%
\column{5}{@{}>{\hspre}l<{\hspost}@{}}%
\column{9}{@{}>{\hspre}l<{\hspost}@{}}%
\column{22}{@{}>{\hspre}l<{\hspost}@{}}%
\column{E}{@{}>{\hspre}l<{\hspost}@{}}%
\>[3]{}\Varid{buildBasicNetwork}{}\<[22]%
\>[22]{}\mathbin{::}\Conid{InitialPipes}\codeskip \Varid{a}\codeskip \Varid{b}{}\<[E]%
\\
\>[22]{}\Rightarrow \Conid{Circuit}\codeskip \Varid{a}\codeskip \Varid{b}\codeskip \Varid{c}\codeskip \Varid{d}\codeskip \Varid{e}\to \Conid{IO}\codeskip (\Conid{BasicNetwork}\codeskip \Varid{a}\codeskip \Varid{b}\codeskip \Varid{c}\codeskip \Varid{d}){}\<[E]%
\\
\>[3]{}\Varid{buildBasicNetwork}\codeskip \Varid{x}\mathrel{=}\mathbf{do}{}\<[E]%
\\
\>[3]{}\hsindent{2}{}\<[5]%
\>[5]{}\Varid{n}{}\<[9]%
\>[9]{}\leftarrow \Varid{icataM}_{5}\codeskip \Varid{buildNetworkAlg}\codeskip \Varid{x}{}\<[E]%
\\
\>[3]{}\hsindent{2}{}\<[5]%
\>[5]{}\Varid{n'}{}\<[9]%
\>[9]{}\leftarrow \Varid{initialNetwork}{}\<[E]%
\\
\>[3]{}\hsindent{2}{}\<[5]%
\>[5]{}\Varid{unAccuN}\codeskip \Varid{n}\codeskip \Varid{n'}{}\<[E]%
\ColumnHook
\end{hscode}\resethooks
\ensuremath{\Varid{icataM}_{5}} builds a value of type \ensuremath{\Conid{AccuN}}, which is unpacked and applied with the \ensuremath{\Varid{initialNetwork}}.
The application step, causes the nest of accumulator functions to collapse and generate a new network, creating the channels and threads.


\end{document}